\documentclass[12pt]{amsart}

\def\centereps#1#2#3{\vskip#2\relax\centerline{\hbox to#1{\special
  {eps:#3 x=#1, y=#2}\hfil}}}
\numberwithin{equation}{section}
\newtheorem{definition}{Definition}[section]

\newtheorem{theorem}{Theorem}[section]
\newtheorem{proposition}{Proposition}[section]
\newtheorem{lemma}{Lemma}[section]
\newtheorem{corollary}{Corollary}[section]
\theoremstyle{remark}
\newtheorem{remark}{Remark}[section]

\newcommand{\loc}{{\text{loc}}}
\newcommand{\vare}{\varepsilon}
\newcommand{\eeps}{(E_{\varepsilon})}

\newcommand{\sca}[2]{\bigl\langle {#1}\ , \; {#2} \bigl\rangle}
\newcommand{\scal}[2]{\bigl\langle {#1} , {#2} \bigl\rangle}

\newcommand{\ep}{\varepsilon}

\newcommand{\al}{\alpha}
\newcommand{\la}{\lambda}

\DeclareMathOperator\dive{div}
\DeclareMathOperator\pf{Pf}
\DeclareMathOperator\curl{curl}

\newcommand{\R}{\mathbb{R}}
\newcommand{\C}{\mathbb{C}}

\begin{document}
\begin{center}
\large\bf
Two dimensional incompressible ideal flow around a small obstacle 
\end{center}
\vskip\baselineskip
\begin{center}
\sc 
D. Iftimie\\
M.C. Lopes Filho\footnote{Research supported in part by CNPq grant \#300.962/91-6}\\
H.J. Nussenzveig Lopes\footnote{Research supported in part by CNPq grant \#300.158/93-9}  
\end{center}
\vskip2\baselineskip
\begin{center}
\parbox{14cm}{\scriptsize
{\sc Abstract.} 
In this article we study the asymptotic behavior of incompressible, ideal, time-dependent two dimensional flow in the exterior of a single smooth obstacle when the size of the obstacle becomes very small. Our main purpose is to identify the equation satisfied by the limit flow. We will see that the asymptotic behavior depends on $\gamma$, the circulation around the obstacle. For smooth flow around a single obstacle, $\gamma$ is a conserved quantity which is determined by the initial data. We will show that if $\gamma = 0$, the limit flow satisfies the standard incompressible Euler equations in the full plane but, if $\gamma \neq 0$, the limit equation acquires an additional forcing term. We treat this problem by first constructing a sequence of approximate solutions to the incompressible 2D Euler equation in the full plane from the exact solutions obtained when solving the equation on the exterior of each obstacle and then passing to the limit on the weak formulation of the equation. We use an explicit treatment of the Green's function of the exterior domain based on conformal maps, {\it a priori} estimates obtained by carefully examining the limiting process and  the Div-Curl Lemma, together with a standard weak convergence treatment of the nonlinearity for the passage to the limit.   
}\end{center}
\vskip1cm

\tableofcontents

\section{Introduction}

        Let $\Omega$ be a bounded, connected and simply connected domain of the plane with smooth boundary $\Gamma$ and, for each $\vare>0$, let $\Omega_{\vare} \equiv \vare \Omega$ with boundary $\Gamma_{\vare}$. We consider a family $u^{\vare}= u^{\vare}(x,t)$ of incompressible flows which satisfy the two-dimensional Euler equations on the exterior of $\Omega_{\vare}$, with velocity tangent to $\Gamma_{\vare}$ and satisfying the initial conditions (i) the initial vorticity $\omega^0(x) = \mbox{ curl }u^{\vare}(x,0)$ is independent of $\vare$ and the support of $\omega_0$ does not intersect the origin and (ii) the  circulation around $\Gamma^{\vare}$ of the velocity $u^{\vare}$ is a real constant $\gamma$ which does not depend on $\vare$ either. Our purpose in this article is to identify the asymptotic behavior when $\vare \to 0$ of the sequence $\{u^{\vare}\}$. We will prove that for $\gamma = 0$, $u^{\vare}$ converges to a flow $u$ satisfying the incompressible two dimensional Euler equations in the full plane with initial vorticity $\omega^0$ and that, for $\gamma \neq 0$, any weak limit of the sequence $\{u^{\vare}\}$ satisfies a modification of the Euler equations which, in the vorticity formulation, takes the aspect of an additional convection term.

In order to clarify the issues involved in this problem, let us consider that, for each fixed time, the velocity $u$ of the flow around a small obstacle decomposed as $u = u_o + u_b$, where $u_b$ is the background velocity, which is slowly varying with respect to the scale of the obstacle, incorporating what we imagine the flow velocity would be if the obstacle was not present, and $u_o$ is the correction to $u_b$ due to the presence of the obstacle. As the obstacle disappears towards a point $p$, the background flow $u_b$ appears as a constant flow background $u_b(p)$ relative to an observer on the obstacle. One can see that, when $u_b(p) \neq 0$, the fact that the full velocity $u$ is tangent to the obstacle means that $u_o$ has to be a large perturbation in absolute value. On the other hand, one expects the perturbation produced by the small obstacle to be sharply localized, so that $|u_o|$ should converge pointwise to zero away from  the obstacle. The small obstacle generates large velocity gradients in a nearby region. How this effect influences the limiting process is the main point of the present work.

It is a well known fact that the ideal flow assumption is physically inappropriate to model the behavior of the flow near an obstacle, due to boundary layer effects. Hence, the whole issue of small obstacle asymptotics would be more physically meaningful if posed for the Navier-Stokes equations. The asymptotics of the ideal flow case which we present here should be regarded as a first step towards the rigorous analysis of the small obstacle problem. 

Most of the work on time-dependent, incompressible exterior flow has been in the nature of well-posedness through energy methods, see \cite{kikuchi83,Ladyzhenskaya69}. Energy estimates are global, so that it would be difficult to treat our sharply localized problem through such means. The alternative is to adopt the vortex dynamics point of view. This means understanding 2D flow in terms of the description of the dynamics of vorticity, an approach which has been very fruitful recently. The most important recent results on 2D Euler are Chemin's Theorem, on the regularity of vortex patches, see \cite{Chemin93}, which is by nature a vortex dynamics result, and Delort's Existence Theorem, see \cite{Delort91}, which relies heavily on the vortex dynamics point of view. 

Our interest in the small obstacle problem was motivated by the problem of confinement of vorticity. Let us examine briefly the nature of this problem and survey some of the results obtained thus far. Consider $\omega = \omega(x,t)$ a (classical) solution of the full plane 2D Euler equations such that $\omega(x,0)$ is compactly supported. The problem of confinement of vorticity is to obtain control over the growth of the diameter of the support of $\omega(\cdot,t)$. Current research on this subject originates with a result obtained by C. Marchioro in \cite{marchioro94}. He proved that the solution of the incompressible 2D Euler equations in the full plane, with bounded, nonnegative initial vorticity with support contained in the ball $B(0;R_0)$, will have, at time $t$, its support contained in a ball of radius $R(t)=(bt + R_0^3)^{1/3}$, for some constant $b \geq 0$.  The state-of-the-art confinement result in the full plane has almost fourth root exponent, see \cite{ISG99,serfati96}. In \cite{marchioro96}, Marchioro addressed the problem of confinement for exterior flow.
Using the techniques developed for full plane flow, Marchioro proved cubic-root confinement for the exterior of a disk and almost square root confinement for a general exterior domain. We will not get into the issue of why the confinement result is sensitive to the presence and to the geometry of the domain, but we observe that confinement is connected to the way an obstacle influences very distant particles, called far-field effects. The scaling behavior of the incompressible Euler equations makes the problem of describing the influence of a very distant obstacle for a very long time naturally associated to observing the effect of a vanishingly small nearby obstacle for some time, which is the object of this article. In this context, the influence of the precise shape of the small domain in the vortex motion is of particular interest.

        From the technical standpoint, this article makes use of the techniques of weak convergence methods for the asymptotic analysis. Such methods have often been developed for proving existence of weak solutions, see for example \cite{Delort91,LNT00}, but they are well suited for studying singular limits in general. The basic ingredients of the present proof are a collection of {\it a priori} estimates obtained mainly through exhaustive use of explicit formulas for the Green's function of the exterior domain and strong compactness of approximate velocities obtained by using the parametrized div-curl Theorem introduced in \cite{LNT00}.  

The remainder of this paper is divided into five sections. In the second section, included mostly for completeness' sake and for fixing notation, we collect information on classical potential theory for the Laplacian in an exterior domain and on holomorphic maps as required to write explicit formulas for the Green's function in terms of the Riemann map associated to the obstacle. In the third section we formulate precisely the exterior flow problem and the problem of small obstacle asymptotics. In the fourth section we derive the collection of {\it a priori} estimates required for the passage to the limit, identifying a collection of quantities that remain under control in the small obstacle asymptotics. We prove our main result in the fifth section, identifying the PDE in the full plane satisfied by the limit flow. In the sixth and last section we collect our conclusions and point the way for future investigation on this problem.

\section{The Laplacian in an exterior domain}

The purpose of this section is to collect a number of facts associated to classical potential theory for the Laplacian on an exterior domain in the plane. We claim no originality on these results, although the explicit form in which they are presented is not the one commonly found in the literature. We include a thorough discussion here for the sake of completeness. 

\subsection{Conformal maps}

Let $\Omega$ be a bounded, open, simply connected subset of the plane, whose boundary, denoted by $\Gamma$, is a $C^{\infty}$ Jordan curve.  We will denote by $\Pi$ the unbounded connected component of $\R^2 - \Gamma$, so that $\Omega^c = \overline{\Pi}$. Let $D = B(0;1)$, $S = \partial D$.

In what follows we identify $\R^2$ with the complex plane $\Bbb{C}$. We begin by constructing a smooth biholomorphism between $\Pi$ and $\{x \in \R^2 \; | \; |x| > 1\} \equiv \mbox{ int } D^c$.

\begin{lemma} \label{confmap} Let $\Pi = \mbox{ int }\Omega^c$ be as above. There exists a smooth biholomorphism $T:\Pi \rightarrow \mbox{ int }D^c$, extending smoothly up to the boundary, mapping $\Gamma$ to $S$. Furthermore, there exists a nonzero real number $\beta$ and a bounded holomorphic function $h: \Pi \to \Bbb{C}$ such that:
\begin{equation} \label{holomform}
T(z) = \beta z + h(z).
\end{equation}
Additionally, 
\begin{equation} \label{morestuff}
h^{\prime}(z) = {\mathcal O}\left(\frac{1}{|z|^2}\right), \mbox{ as } |z| \to \infty.
\end{equation}
\end{lemma}
\begin{proof}
We may assume without loss of generality that $B(0;1) \subset \Omega$. If this were not the case then suitable dilation and translation would make it so.

For $z \in \Pi$, consider the  holomorphic map given by $z \mapsto I(z) = 1/z$ and denote by $\widetilde{\Omega}$, $\widetilde{\Gamma}$, the image of $\Pi$, $\Gamma$  respectively, under this map. There exists a biholomorphism ${\mathcal R}$ from $\widetilde{\Omega}\cup\{0\}$ to $B(0;1)$ given by the Riemann Mapping Theorem, mapping $\widetilde{\Gamma}$ to $S$. Since $\Gamma$ was assumed to be a $C^{\infty}$ Jordan curve so is $\widetilde{\Gamma}$, hence ${\mathcal R}$ and its inverse ${\mathcal R}^{-1}$ have continuous extensions up to the boundary, along with all their derivatives (see \cite{BK85}). 
We choose the Riemann mapping ${\mathcal R}$ such that ${\mathcal R}(0) = 0$ and we define $T$ to be given by the compositions: $T = I \circ {\mathcal R} \circ I$. It is easy to see that $T$ is a biholomorphism, continuous up to $\Gamma$, along with all of its derivatives. 

Consider the function $f$ defined by 
\[ {\mathcal R} (z) = zf(z), \; z \in \widetilde{\Omega}. \]
This is a holomorphic function on $\widetilde{\Omega}$ which cannot vanish at $0$ since ${\mathcal R}$ is a bijection. By the same token, $f$ cannot vanish anywhere else in the closure of $\widetilde{\Omega}$. It follows that 
 $1/f(z)$ is also a holomorphic function. Hence we can write $1/f(z) = \beta + zg(z)$, with $g$ a holomorphic function and $\beta = 1/ f(0)$. We find $g(z) = -(f(z) - f(0))/f(0){\mathcal R}(z)$, which can be seen to be bounded since $z=0$ is a zero of order $1$ for ${\mathcal R}$ and $f$ is bounded. 

Thus we obtain
\[ T(z) = 1/{\mathcal R}(1/z) = z(\beta + (1/z) g(1/z)) = \beta z + g(1/z).\]
The desired function $h$ is given by $h(z) = g(1/z)$. 
We may assume without loss of generality that $\beta$ is real by multiplying the function $T$ constructed above by $\overline{\beta}/|\beta|$ if necessary, which does not change the desired properties.
Finally, we observe that 
\[ h^{\prime}(z) = -\frac{1}{z^2}g^{\prime}\left(\frac{1}{z}\right), \]
and, by construction, $g^{\prime}$ is a holomorphic function in $\widetilde{\Omega}$, bounded in the closure.  
\end{proof}

\begin{remark} It follows by construction of $T$ that 
\[T^{\prime}(z) = {\mathcal R}^{\prime}(1/z)/(f(1/z))^2,\]
which can be easily seen to be bounded from above and below. Therefore, if we make the canonical identification of $\R^2$ with $\Bbb{C}$, writing $z = x_1 + i x_2$ for the point $x=(x_1,x_2)$, then, if $DT$ stands for the Jacobian matrix associated to $T^{\prime}$, there exist a positive constant $C$ such that 
\begin{equation} \label{derivest}
\| DT \|_{L^{\infty}} \leq C \mbox{ and } \| DT^{-1} \|_{L^{\infty}} \leq C.
\end{equation}
\end{remark}

\subsection{Explicit formulas for the Green's function}

Here we will obtain an explicit formula for the Green's function of the Laplacian in $\Pi$ in terms of the conformal mapping above. We denote this Green's function by $G_{\Pi}=G_{\Pi}(x,y)$; we must have $\Delta_y G_{\Pi}(x,y) = \delta(y-x)$, $G_{\Pi}(x,y) = 0$ for $y \in \Gamma$ and $G_{\Pi}(x,y) = G_{\Pi}(y,x)$. If $\Omega = B(0;1)$ and if we write $x^{\ast} = x/|x|^2$ then there is a unique Green's function which can be written explicitly as
\[ G_{D^c}(x,y) =  \frac{1}{2\pi}\log \frac{|x-y|}{|x-y^{\ast}||y|}. \]
 
It is easy to verify that if $x_0 \in (B(0;1))^c$ and $h$ satisfies $\Delta h = \delta(x-x_0)$ in a neighborhood of $x_0$ then $\tilde{h} \equiv h \circ T$ satisfies $\Delta \tilde{h} = \delta(y-T^{-1}(x_0))$ in a neighborhood of $T^{-1}(x_0)$. We use this fact on $G$ to write:
\begin{equation} \label{greensfctn}
G_{\Pi}(x,y) = \frac{1}{2\pi}\log \frac{|T(x)-T(y)|}{|T(x)-(T(y))^{\ast}||T(y)|}. 
\end{equation}

We will concern ourselves mostly with first order derivatives of the Green's function, which we will introduce through the notation $K_{\Pi} = K_{\Pi}(x,y) \equiv \nabla^{\perp}_x  G_{\Pi}(x,y)$. The explicit formula for $K_{\Pi}$ is given by
\begin{equation} \label{kpidef}
K_{\Pi}(x,y) = \frac{((T(x) - T(y))DT(x))^{\perp}}{2\pi|T(x) - T(y)|^2} - 
\frac{((T(x) - (T(y))^{\ast})DT(x))^{\perp}}{2\pi|T(x) - (T(y))^{\ast}|^2}.
\end{equation}
Note that $K_{\Pi}(x,y) = DT^t(x) K_{D^c}(T(x),T(y))$.

We require information on far-field behavior of $K_{\Pi}$. We will use
several times the following general relation:  
\begin{equation} \label{drid}
  \Bigl|\frac{a}{|a|^2}-\frac{b}{|b|^2}\Bigr|
= \frac{|a-b|}{|a||b|}
\end{equation}
which can be readily checked by squaring both sides.

We now find from the estimates (\ref{derivest}), \eqref{kpidef} and (\ref{drid}), that 
\[|K_{\Pi}(x,y)| \leq C\frac{|T(y)-(T(y))^{\ast}|}{|T(x)-T(y)||T(x)-(T(y))^{\ast}|}.\]
For $f \in C^{\infty}_c(\Pi)$, we introduce the notation
\begin{equation} \label{notop}
K_{\Pi}[f] = K_{\Pi}[f](x) \equiv \int_{\Pi} K_{\Pi}(x,y)f(y)dy. 
\end{equation}

It is easy to see that the pointwise estimate for $K_{\Pi}$ above, yields, for large $|x|$, the estimate:
\begin{equation} \label{inftyworry}
|K_{\Pi}[f]|(x) \leq \frac{C}{|x|^2}, 
\end{equation}
where the constant $C>0$ depends on the size of the support of $f$. In the last inequality we have used the explicit formula for the biholomorphism $T$ (\ref{holomform}).

\begin{lemma} \label{ellipobs}
The vector field $u = K_{\Pi}[f]$ is a solution to the elliptic system:
\[ \left\{ \begin{array}{l}
         \mbox{div }u = 0       \mbox{ in } \Pi \\
         \mbox{curl }u = f \mbox{ in } \Pi  \\
         u \cdot \hat{n} = 0 \mbox{ on } \Gamma \\
         \lim_{|x| \to \infty}|u| = 0.
\end{array} \right. \]
\end{lemma}

The proof of this Lemma is straightforward.
 
\subsection{Harmonic Vector Fields}

 The standard version of Hodge's Theorem is proved for compact manifolds, see \cite{warner71}, with a natural extension to compact manifolds with boundary. Extending Hodge's Theory to noncompact manifolds is a difficult problem, see \cite{lockhart87} for a broad overview. There is a lot of special structure for the exterior domain, specifically the complex structure, which allows us to prove an elementary extension of Hodge's Theorem to this particular case. It is no surprise that this extension is valid, moreover, such a fact has been extensively used in the literature. As before, we include a complete discussion here because we will require the very explicit treatment involved for the remainder of this article.

  Let $\Omega$ be a bounded, open subset of $\R^2$ whose boundary $\Gamma$ is a smooth Jordan curve and let $\Pi = \mbox{ int } \Omega^c$. We will denote by $\hat{n}$ the unit normal {\it exterior} to $\Omega^c$ at $\Gamma$. In what follows {\it all contour integrals} are taken in the {\it counter-clockwise} sense, so that $\int_{\Gamma} F \cdot \mathbf{ds} = - \int_{\Gamma} F \cdot \hat{n}^{\perp} ds$.

\begin{proposition} \label{anality}
There exists a unique classical solution $H = H_{\Pi}$ of the problem
\begin{equation} \label{harmvecfields}
    \left\{ \begin{array}{l}
                \mbox{div }H = 0,  \mbox{ in } \Pi, \\
                \mbox{curl }H = 0,  \mbox{ in } \Pi, \\
                H \cdot \hat{n} = 0 \mbox{ on } \Gamma, \\
                |H| \to 0 \mbox{ as } x \to \infty, \\
                \int_{\Gamma} H \cdot \mathbf{ds} = 1. \end{array} \right. \end{equation}
Moreover, $H_{\Pi} = {\mathcal O}(1/|x|)$ when $|x| \to \infty$.
\end{proposition}   
  
\begin{proof}
We will start by proving uniqueness. Let us assume there are two solutions $H_1, \; H_2$ and consider their difference $H = H_1 - H_2$. Since the problem is linear $H$ is a harmonic vector field with zero circulation. We must show that $H$ vanishes identically. Let us begin by estimating $H$ at infinity. We will use the canonical identification of $\R^2$ with $\C$, writing $z = x_1 + i x_2$ instead of $x = (x_1, x_2)$. First note that $H = O(|x|^{-1})$ as $x \to \infty$. To see this consider the holomorphic function $f = f(z) = \bar{H}(1/z)$. The singularity 
at $z=0$ is clearly removable and hence $f$ is holomorphic in a neighborhood of the origin. Furthermore $f(0) = 0$, which implies that $f(z)/z$ is also holomorphic in a neighborhood of $0$. Thus we have shown the desired behavior at infinity. Next, note that, after an easy calculation we find, for any real vector field $L$,
\begin{equation} \label{magicoint}
 \oint_{\Gamma} \bar{L}dz = \int_{\Gamma} L \cdot \mathbf{ds} - i \int_{\Gamma} L \cdot \hat{n} ds.
\end{equation}
Using this identity for $H$ we conclude that the contour integral of $\bar{H}$ on $\Gamma$ vanishes. Since $\bar{H}$ is holomorphic in $\Omega^c$, its contour integral on any closed curve enclosing $\Omega$ vanishes as well. Let us consider the Taylor expansion of $f=f(z)$ at $z = 0$, $f(z) = cz + O(z^2)$. Fix $\vare > 0$ suitably small. We have that:
\[ 2\pi i c = \oint_{|z| = \vare} \frac{f(z)}{z^2}dz = \oint_{|w| = \vare^{-1}} \bar{H}(w)dw = 0.\]
From this we conclude that actually $H = O(|x|^{-2})$ as $x \to \infty$. 

Next we observe that, since $H$ is curl-free and has zero circulation, there exists a function $\varphi$ such that $H = \nabla \varphi$. This function is constructed by integrating along arbitrary paths. From the behavior of $H$ at infinity and the construction of $\varphi$ we have that $|\varphi(x)| = O(|x|^{-1})$ as $x \to \infty$. 
 
Let us now use the estimates gathered above to show that $H \equiv 0$. We integrate:
\begin{equation*} 
\int_{\Omega^c} |H|^2 dx = \lim_{R \to \infty} \int_{B(0;R)\setminus \Omega} |H|^2 dx 
=\lim_{R \to \infty} \int_{B(0;R)\setminus \Omega} H \cdot \nabla \varphi dx = 
 \lim_{R \to \infty} \int_{|x| = R} \varphi H \cdot \frac{x}{R} ds, 
\end{equation*}
by integration by parts and using the fact that $H$ is tangent to $\Gamma$. This limit vanishes because of the behavior of $\varphi$ and $H$ at infinity. Thus we conclude that $H$ is identically zero in $\Omega^c$.  

We now obtain the existence of a classical solution. 
Let $T$ be the biholomorphic mapping from Lemma \ref{confmap}.   
Let $\psi = \psi(z) = \log | T(z) |$ for $z \in \Omega^c$.
Since the logarithm is a harmonic function in $B(0;1)^c$ and $T$ is analytic in $\Pi$ it follows that $\psi$ is a harmonic function of $x$. Furthermore $\psi \equiv 0$ on $\Gamma$. Define $U \equiv \nabla^{\perp}\psi = (-\partial_{x_2}\psi, \partial_{x_1}\psi)$. It can be easily checked that $U$ satisfies all but one of the conditions in system (\ref{harmvecfields}), namely the condition on circulation on $\Gamma$. Since $U$ is not identically zero it follows from the uniqueness part of the proof that
\[\int_{\Gamma} U \cdot \mathbf{ds} = c \neq 0,\]
and hence we may take $H_{\Pi} = U/c$. 

Finally we address the asymptotic behavior of $H_{\Pi}$ at infinity. Recall that, in the beginning of the proof of uniqueness above, we showed that the difference of two solutions behaved like $O(|x|^{-1})$ as $x \to \infty$. The argument we gave did not depend on the circulation of the difference $H$. Therefore the same argument can be used with $H_1 = H_{\Pi}$ and $H_2 = 0$ to show the desired behavior for $H_{\Pi}$.
\end{proof}

By construction the harmonic vector field $H_{\Pi}$ above is given by 
\[ H_{\Pi}(x) = C\nabla^{\perp} \log |T(x)|,\]
for some constant $C$. We argue that $C = 1/2\pi$. Indeed, it is an easy calculation to prove that
\[ \bar{H}_{\Pi}(z) = -Ci (\log  T)^{\prime}(z) = -Ci \frac{T^{\prime}(z)}{T(z)}.\]
Using (\ref{magicoint}) together with the fact that $H_{\Pi}$ is tangent to $\Gamma$ we find:
\[ \int_{\Gamma} H_{\Pi} \cdot \mathbf{ds} = \oint_{\Gamma} \bar{H}_{\Pi} dz = C\oint_{\Gamma}
 \frac{T^{\prime}(z)}{iT(z)}dz = C \oint_{S} \frac{dw}{iw} = 2\pi C,\]
so that we must have $C = 1/2\pi$ in order to satisfy the last condition of system (\ref{harmvecfields}). Throughout the remainder of this paper we set 
\begin{equation} \label{theH}
H_{\Pi}(x) = \frac{1}{2\pi} \nabla^{\perp} \log |T(x)| = \frac{1}{2\pi} \frac{(T(x)DT(x))^{\perp}}{|T(x)|^2} =  \frac{1}{2\pi} \frac{DT^t(x)(T(x))^{\perp}}{|T(x)|^2}.
\end{equation}

%Finally, from the explicit formula for the harmonic vector field we obtain:
%\begin{equation} \label{Hcomment}
%H_{\Pi} \in L^p(\Pi) \mbox{ for any } p > 2.
%\end{equation}
  
\section{Flow in an exterior domain}

The purpose of this section is to formulate precisely the small obstacle limit.  

\subsection{The initial-boundary value problem}
         We begin by formulating precisely the initial-boundary value problem for incompressible ideal fluid flow in an exterior domain. Let $\Gamma$ be, as before, a smooth Jordan curve in the plane, dividing it into a bounded connected component, which we call $\Omega$ and an unbounded connected component denoted $\Pi$. For $x \in \Gamma$, denote by $\hat{n}(x)$ the exterior normal to $\Omega^c$ at $x$, as before.  

Let $u = u(x,t) = (u_1(x_1,x_2,t),u_2(x_1,x_2,t))$ be the velocity of an incompressible, ideal fluid in $\Omega^c$. We assume that $u$ is tangent to $\Gamma$ and $u \to 0$ when $|x| \to \infty$. The evolution of such a flow is governed by the Euler equations. We write the initial-boundary value problem as:

\begin{equation} \label{eulereq}
\left\{ \begin{array}{ll}
u_t + u \cdot \nabla u = - \nabla p & \mbox{ in } \Pi \times (0,\infty)  \\
\mbox{div }u = 0 & \mbox{ in } \Pi \times [0,\infty) \\
u \cdot \hat{n} = 0 & \mbox{ on } \Gamma \times [0,\infty) \\
\lim_{|x| \to \infty}|u| = 0 & \mbox{ for } t \in [0,\infty)\\
u(x,0) = u_0(x) & \mbox{ in } \Omega^c,
\end{array} \right. \end{equation} 
where $p = p(x,t)$ is the scalar pressure. If $u_0$ is sufficiently smooth, global well-posedness of this problem was proved by K. Kikuchi in \cite{kikuchi83}. 

Let $\omega = \mbox{curl }u$ be the vorticity associated to this flow. In order to write a vorticity formulation of problem (\ref{eulereq}) we must be able to recover velocity from vorticity. The coupling of velocity and vorticity is given by the elliptic system
\[\left\{ \begin{array}{l}
         \mbox{div }u = 0       \mbox{ in } \Pi \times [0,\infty) \\
         \mbox{curl }u = \omega \mbox{ in } \Pi \times [0,\infty) \\
         u \cdot \hat{n} = 0 \mbox{ on } \Gamma \times [0,\infty) \\
         \lim_{|x| \to \infty}|u| = 0 \mbox{ for } t \in [0,\infty).
\end{array} \right. \]

In view of Lemma \ref{ellipobs} and Proposition \ref{anality} this system  
has a unique solution up to a harmonic vector field, given by $u = u(x,t) =  K_{\Pi}[\omega(\cdot,t)](x) + \alpha H_{\Pi}(x)$, for some 
time-dependent function $\alpha = \alpha(t)$. 

\begin{lemma} \label{alphaconst} 
If $u$ is a smooth solution of (\ref{eulereq}) then $\alpha$ is  constant in time. 
\end{lemma}

\begin{proof}
We introduce the stream function $\psi = \psi(x,t)$ given by:
\begin{equation} \label{strfnct}
\psi (x,t) \equiv G_{\Pi}[\omega](x,t) = \int_{\Pi} G_{\Pi}(x,y)\omega(y,t)dy.
\end{equation}

Next, we observe that, by Kelvin's Circulation Theorem, the circulation of $u$ around $\Gamma$ is constant in time. Hence we have:
\[ \gamma = \int_{\Gamma} u \cdot {\mathbf ds} = \int_{\Gamma} \nabla^{\perp} \psi \cdot {\mathbf ds} + \alpha(t) \int_{\Gamma}H_{\Pi} \cdot {\mathbf ds}
= \alpha(t) - \int_{\Gamma} \nabla \psi \cdot \hat{n} ds. \]
Integrating by parts we find:
\[\gamma = \alpha(t) - \lim_{R \to \infty}\left( \int_{B(0;R)\setminus \Omega} \omega dx - \int_{\partial B(0;R)} \nabla \psi \cdot \frac{x}{R} ds \right)
= \alpha(t) - \int_{\Pi}\omega dx, \]
using (\ref{inftyworry}).
Hence, since mass of vorticity is conserved,
\[ \alpha(t) \equiv \gamma + \int_{\Pi} \mbox{ curl }u_0 dx. \] 
\end{proof} 

Finally, we can now write the vorticity formulation of this problem as:
\begin{equation} \label{vorteq}
\left\{ \begin{array}{ll}
\omega_t + u \cdot \nabla \omega = 0, & \mbox{ in } \Pi \times (0,\infty)  \\
u = K_{\Pi}[\omega] + \alpha H_{\Pi}, & \mbox{ in } \Pi \times [0,\infty)\\
\omega(x,0) = \mbox{curl }u_0 (x), & \mbox{ in } \Pi. 
\end{array} \right. \end{equation}

\subsection{The evanescent obstacle} 

In this subsection we will formulate a family of problems, parametrized by the size of the obstacle, in order to identify the asymptotic limit under consideration. 
Fix $\omega_0 = \omega_0(x) \in C^{\infty}_c(\R^2)$ and assume that the origin does not belong to the support of $\omega_0$. Let $\Omega$ be a domain in the plane satisfying the hypothesis of Lemma \ref{confmap}. We will consider the family of rescaled domains $\Omega_{\vare} \equiv \vare \Omega$ and we note that there exists $\vare_0$ such that, for all $0< \vare \leq \vare_0$, the support of $\omega_0$ does not intersect $\Omega_{\vare}$. 

Let $\Gamma_{\vare} = \partial \Omega_{\vare}$ and $\Pi_{\vare} = \mbox{ int } \Omega^c_{\vare}$. We denote the harmonic vector field given by Proposition \ref{anality} by $H^{\vare} = H_{\Pi_{\vare}}$. We also denote the Green's function for $\Pi_{\vare}$ by $G^{\vare}$ and the corresponding kernel (and integral operator) $K^{\vare} = \nabla^{\perp}G^{\vare}$.

Consider the system $\eeps$ given below:
\[ \eeps \left\{ \begin{array}{ll}
\omega^{\vare}_t + u^{\vare} \cdot \nabla \omega^{\vare} = 0, & \mbox{ in } \Pi_{\vare} \times (0,\infty)  \\
u^{\vare} = K^{\vare}[\omega^{\vare}] + \alpha H^{\vare}, & \mbox{ in } \Pi_{\vare} \times [0,\infty)\\
\omega^{\vare}(x,0) = \omega_0 (x), & \mbox{ in } \Pi_{\vare}. 
\end{array} \right. \]
It follows from the work of Kikuchi \cite{kikuchi83} that, for any $\vare > 0$, if $\omega_0$ is sufficiently smooth then this system has a unique smooth solution. 

In fact, Kikuchi's result asserts existence for the velocity formulation of $\eeps$, which means that there exists a pressure $p^{\vare}$ such that $u^{\vare}$ and $p^{\vare}$ are a solution of problem (\ref{eulereq}) in the domain $\Pi_{\vare}$ with initial velocity 
\begin{equation} \label{velinitvare}
u^{\vare}_0 = K^{\vare}[\omega_0] + \alpha H^{\vare}. 
\end{equation}

Consider $T$ the biholomorphism from $\Pi$ to $\mbox{ int } D^c$ constructed in Lemma \ref{confmap}. Observe that $T^{\vare}(z) \equiv T(z/\vare)$ defines a biholomorphism between $\Pi_{\vare}$ and  $\mbox{ int } D^c$, which extends smoothly up to the boundary and which maps $\Gamma_{\vare}$ to $S$. We can use $T^{\vare}$ to write explicit formulas for $K^{\vare}$ and $H^{\vare}$, by recalling (\ref{kpidef}) and (\ref{theH}). We have
\begin{equation}\label{kpivare}
\begin{split} 
K^{\vare} 
&= \frac{1}{\vare}
K_{\Pi}\left(\frac{x}{\vare},\frac{y}{\vare}\right)\\
&=\frac{1}{2\pi\vare}DT^t(x/\vare) \left(\frac{((T(x/\vare) - T(y/\vare)))^{\perp}}{|T(x/\vare) - T(y/\vare)|^2} - 
\frac{((T(x/\vare) - (T(y/\vare))^{\ast}))^{\perp}}{|T(x/\vare) - (T(y/\vare))^{\ast}|^2} \right),
\end{split}
\end{equation}
and
\begin{equation} \label{hvare}
H^{\vare} = \frac{1}{\vare} H_{\Pi}\left(\frac{x}{\vare}\right) = 
\frac{1}{2\pi\vare}DT^t(x/\vare)\left( \frac{(T(x/\vare))^{\perp}}{|T(x/\vare)|^2}\right). 
\end{equation}

We will require information on the behavior  of $H^{\vare}$ as $\vare \to 0$. One easy observation on that regard is
the fact that for any $R>0$, there exists $C = C(R) > 0$ such that 
\begin{equation} \label{cizania}
\|H^{\vare}\|_{L^1(\Pi_{\vare} \cap B(0;R))} \leq C,
\end{equation} 
uniformly in $\vare$. Indeed,
\begin{align*}
\int_{\Pi_{\vare} \cap B(0;R)} |H^{\vare}|dx 
&= \frac{C}{\vare}\int_{\Pi_{\vare} \cap B(0;R)} \left| \frac{(DT^t(x/\vare)(T(x/\vare))^{\perp})}{|T(x/\vare)|^2}\right| dx \\
&\leq \int_{1 \leq |y| \leq C(R)/\vare} \frac{C}{\vare} \frac{1}{|y|}
\vare^2 |\det (DT^{-1})(y)|dy, 
\end{align*}
where we used the change of variables $y = T(x/\vare)$, we used (\ref{derivest}) and $C(R)$ is a suitable 
constant computed using the expression for $T$ from Lemma \ref{confmap}. From this last expression, estimate (\ref{cizania}) follows easily.

\section{A priori estimates}

        In this section we will prove the a priori estimates on $u^{\vare}$ and $\omega^{\vare}$ which are required to identify their asymptotic behavior.

\subsection{Velocity estimate}

        The key ingredients on the rigorous treatment of weak solutions for the incompressible Euler equations are usually the energy estimate on velocity and estimates on vorticity based on the fact that the vorticity is rearranged by incompressible flow, i.e. the distribution function of vorticity is a constant of motion. There are numerous instances of this observation, and we refer the reader to \cite{LNT00} for a systematic discussion. For the problem in hand, the   
usual rearrangement estimates on vorticity hold for the sequence $\omega^{\vare}$, due to the transport nature of equation $\eeps$. For $\vare$ small enough, so that the support of $\omega_0$ is compactly contained on $\Pi_{\vare}$ we have the a priori bounds
\begin{equation} \label{vortest}
\|\omega^{\vare}(\cdot,t)\|_{L^{\infty}(\Pi_{\vare})} = \|\omega_0\|_{L^{\infty}(\R^2)}
\mbox{ and } \|\omega^{\vare}(\cdot,t)\|_{L^1(\Pi_{\vare})} = \|\omega_0\|_{L^1(\R^2)}
\end{equation}

        On the other hand, we are not able to use the energy estimate in the usual way. In order to explain what is at play here, let us illustrate the behavior of the velocity as $\vare \to 0$ with an explicit example.

\vspace{0.5cm}

{\bf Example:} Let $\Omega_{\vare} = B(0;\vare)$, so that $\Pi_{\vare} = \{|x|>\vare\}$. We consider the vorticity 
\[ \omega = \omega(x) \equiv \left\{ \begin{array}{l} 1 \mbox{ if } 1 \leq |x| \leq 2 \\
0 \mbox{ otherwise,} \end{array} \right. \]
which is a stationary solution of $\eeps$ for any $\vare < 1$.

The finite energy part of the associated velocity is $v^{\vare} \equiv K^{\vare}[\omega]$, which, after straightforward calculations gives
\[ v^{\vare}(x) =  \left\{ \begin{array}{ll} -\frac{3}{2} \frac{x^{\perp}}{|x|^2} & \mbox{ if } \vare \leq |x| < 1 \\ \\
-\frac{3}{2} \frac{x^{\perp}}{|x|^2} + \frac{x^{\perp}}{2} \left( \frac{|x|^2-1}{|x|^2} \right)
& \mbox{ if } 1 \leq |x| < 2 \\ \\
0 & \mbox{ if } |x| \geq 2 \end{array} \right. \]   

Clearly, there is (logarithmic) local blow-up of the kinetic energy as $\vare \to 0$. The harmonic vector field associated to $\Pi_{\vare}$ is the restriction of $H(x) = x^{\perp}/2\pi |x|^2$ to $\Pi_{\vare}$, so that the full velocity is $u^{\vare} = v^{\vare} + \alpha H$. If we want to single out a locally uniformly square integrable portion of $u^{\vare}$ for estimating we must add a well chosen part of $\alpha H$ to $v^{\vare}$. We consider $u^{\vare} = (v^{\vare} + 3\pi H) + (\alpha-3\pi)H$. Neither part of this decomposition is square-integrable at infinity, but $v^{\vare} + 3\pi H$ is not only uniformly bounded in $L^2_{\loc}$ but it is actually uniformly bounded in $L^{\infty}$. We observe that $3\pi = \int \omega (x) dx$. The purpose of this subsection is to prove that something similar holds in general.

\vspace{0.5cm}

Returning to the general sequence $\omega^{\vare},u^{\vare}$, assume again that $\vare$ is small enough so that the support of $\omega_0$ is compactly contained in $\Pi_{\vare}$. We introduce \[ m \equiv \int_{\Pi_{\vare}} \omega^{\vare} dx = \int_{\R^2} \omega_0 dx. \]
Introduce also
\[v^{\vare} \equiv K^{\vare}[\omega^{\vare}] + m H^{\vare}, \]
so that $u^{\vare} = v^{\vare} + (\alpha-m) H^{\vare}$.

\begin{theorem} \label{supest}
There exists a constant $C>0$ that depends only on the shape of $\Omega$ such that 
\begin{equation*} %\label{thingb}
\|v^{\vare}\|_{L^{\infty}(\Pi_{\vare})} \leq C\|\omega^{\vare}\|_{L^{\infty}}^{1/2} \|\omega^{\vare}\|_{L^1}^{1/2}.
\end{equation*}
\end{theorem}

\begin{proof}
We write $v^{\vare}(x,t) = {\mathcal I}_1 + {\mathcal I}_2$ explicitly using (\ref{kpivare}) and (\ref{hvare}) with
\[{\mathcal I}_1 = \frac{1}{2\pi\vare}DT^t(x/\vare) \int_{\Pi_{\vare}}\frac{(T(x/\vare) - T(y/\vare))^{\perp}}{|T(x/\vare) - T(y/\vare)|^2} \omega^{\vare}(y,t) dy, \]
and
\[{\mathcal I}_2 = \frac{1}{2\pi\vare}DT^t(x/\vare)  \int_{\Pi_{\vare}} \left(
-\frac{(T(x/\vare) - (T(y/\vare))^{\ast})^{\perp}}{|T(x/\vare) - (T(y/\vare))^{\ast}|^2}
+ \frac{T(x/\vare)^{\perp}}{|T(x/\vare)|^2}\right) \omega^{\vare}(y,t) dy. \]

We begin by estimating ${\mathcal I}_1$. Using (\ref{derivest}) we get

\[|{\mathcal I}_1| \leq \frac{C}{\vare} \int_{\Pi_{\vare}}\frac{1}{|T(x/\vare) - T(y/\vare)|}
|\omega^{\vare}|(y,t)dy. \]

Let $J = J(\xi) \equiv |\det(DT^{-1})(\xi)|$ and $z = \vare T(x/\vare)$. Then, making the change of variables $\eta = \vare T(y/\vare)$, we find
\begin{equation}\label{i1}
|{\mathcal I}_1| \leq C \int_{|\eta|\geq \vare} \frac{1}{|z-\eta|}
 |\omega^{\vare}(\vare T^{-1}(\eta/\vare),t)| J(\eta/\vare) d\eta. 
\end{equation}

Next, we introduce 
\begin{equation} \label{effepsilon}
f^{\vare}(\eta,t) = |\omega^{\vare}(\vare T^{-1}(\eta/\vare),t)| J(\eta/\vare)\chi_{|\eta|\geq \vare}, 
\end{equation}
where $\chi_E$ is the characteristic function of the set $E$. We change variables back and we get:
\begin{equation} \label{eff1}
\|f^{\vare}(\cdot,t)\|_{L^1(\R^2)} = \|\omega^{\vare}(\cdot,t)\|_{L^1(\R^2)} , 
\end{equation}
and, since by (\ref{derivest}) $J$ is bounded, we find that
\begin{equation} \label{effinfty}
\|f^{\vare}(\cdot,t)\|_{L^{\infty}(\R^2)} \leq C \|\omega^\vare\|_{L^{\infty}}. 
\end{equation}

We deduce from \eqref{i1} and \eqref{effepsilon} that
\begin{equation} \label{ai1est}
|{\mathcal I}_1| \leq C\int_{\R^2} \frac{1}{|z-\eta|} f^{\vare}(\eta,t) d\eta \leq
C \|f^{\vare}\|_{L^1}^{1/2} \|f^{\vare}\|_{L^{\infty}}^{1/2}. 
\end{equation}
This last inequality is Lemma 2.1 in \cite{ISG99}. According to
\eqref{eff1} and \eqref{effinfty} this concludes the estimate for
${\mathcal I}_1$.

Let us now estimate ${\mathcal I}_2$. Using again (\ref{derivest}) we find

\[ |{\mathcal I}_2| \leq \frac{C}{\vare} \int_{\Pi_{\vare}} \left|
\frac{T(x/\vare) - (T(y/\vare))^{\ast} }{|T(x/\vare) - (T(y/\vare))^{\ast}|^2}
- \frac{T(x/\vare)}{|T(x/\vare)|^2}\right| |\omega^{\vare}(y,t)| dy.\]

Again we consider $J = J(\xi) \equiv |\det(DT^{-1})(\xi)|$ and $z = \vare T(x/\vare)$ and we make the same change of variables $\eta = \vare T(y/\vare)$. Using (\ref{drid}) this yields
\begin{align*}
|{\mathcal I}_2| &\leq \frac{C}{\vare} \int_{|\eta| \geq \vare} \left| \frac{z/\vare - \vare \eta^{\ast}}{|z/\vare - \vare \eta^{\ast}|^2} - \frac{z/\vare}{|z/\vare|^2}\right|
|\omega^{\vare}(\vare T^{-1}(\eta/\vare),t)|J(\eta/\vare)d\eta \\
&\leq C \int_{|\eta|\geq \vare} \frac{\vare^2 |\eta^{\ast}|}{|z||z-\vare^2\eta^{\ast}|} 
|\omega^{\vare}(\vare T^{-1}(\eta/\vare),t)|J(\eta/\vare)d\eta. 
\end{align*}

As $z = \vare T(x/\vare)$ and the image of $T$ is the exterior of the unit disk, it follows that $|z|\geq\vare$. Hence,
\[|{\mathcal I}_2| \leq C \int_{|\eta|\geq \vare} \frac{\vare |\eta^{\ast}|}{|z-\vare^2\eta^{\ast}|} 
|\omega^{\vare}(\vare T^{-1}(\eta/\vare),t)|J(\eta/\vare)d\eta. \]

We again change variables in the integral above, writing $\theta = \vare \eta^{\ast}$. Then we have:
\[|{\mathcal I}_2| \leq C \int_{|\theta| \leq 1} \frac{|\theta|}{|z-\vare\theta|} |\omega^{\vare}(\vare T^{-1}(\theta^{\ast}),t)|J(\theta^{\ast}) \frac{\vare^2}{|\theta|^4} d\theta \]
\[ = C \left(\int_{|\theta| \leq 1/2} + \int_{1/2 \leq |\theta| \leq 1}\right) \equiv C({\mathcal I}_{21} + {\mathcal I}_{22}). \] 

First we estimate ${\mathcal I}_{21}$. For $|\theta| \leq 1/2$ we have that $|z - \vare \theta| \geq \vare/2$, so that
\begin{multline*}
|{\mathcal I}_{21}| \leq \int_{|\theta|\leq 1/2} 2\vare |\theta||\omega^{\vare}(\vare T^{-1}(\theta^{\ast}),t)|J(\theta^{\ast}) \frac{d\theta}{|\theta|^4} = \\
= 2 \int_{|\eta| \geq 2\vare} \frac{|\omega^{\vare}(\vare T^{-1}(\eta/\vare),t)|J(\eta/\vare)}{|\eta|}d\eta \leq
2 \int_{\R^2} \frac{f^{\vare}(\eta,t)}{|\eta|}d\eta,
\end{multline*}
where $f^{\vare}$ was introduced in (\ref{effepsilon}).
Hence, it follows from (\ref{eff1}), (\ref{effinfty}) and (\ref{ai1est}) with $z=0$ that $|{\mathcal I}_{21}| \leq C\|\omega^\vare\|_{L^1}^{1/2}  \|\omega^\vare\|_{L^\infty}^{1/2}$.

Finally, we estimate ${\mathcal I}_{22}$. Let 
\begin{equation*}
g^\vare(\theta,t)=|\omega^{\vare}(\vare T^{-1}(\theta^{\ast}),t)|J(\theta^{\ast}) \frac{\vare^2}{|\theta|^4}  
\end{equation*}
so that
\begin{equation}\label{dragos}
{\mathcal I}_{22}=\int_{1/2 \leq |\theta| \leq 1}\frac{|\theta|}{|z-\vare\theta|}g^\vare(\theta,t)d\theta.   
\end{equation}
As above, we deduce by changing variables back that 
\begin{equation} \label{dragos1}
\|g^\vare\|_{L^1(1/2 \leq |\theta| \leq 1)}\leq \|\omega^\vare\|_{L^1}. 
\end{equation}
Also, it is trivial to see that
\begin{equation} \label{dragos2}
\|g^\vare\|_{L^\infty(1/2 \leq |\theta| \leq 1)}\leq \vare^2 \|\omega^\vare\|_{L^\infty}.  
\end{equation}
By factoring out $\vare$ in the denominator we can re-write (\ref{dragos}) and use Lemma 2.1 in \cite{ISG99} to deduce that:
\[ |{\mathcal I}_{22}| = \frac{1}{\vare} 
\int_{1/2 \leq |\theta| \leq 1} 
\frac{|\theta|}{|(z/ \vare)-\theta|}g^\vare(\theta,t)d\theta \leq \frac{C}{\vare}  
\|g^\vare\|_{L^1(1/2 \leq |\theta| \leq 1)}^{1/2}  
\|g^\vare\|_{L^{\infty}(1/2 \leq |\theta| \leq 1)}^{1/2} \]
\[\leq C \|\omega^{\vare}\|_{L^1}^{1/2}  \|\omega^{\vare}\|_{L^\infty}^{1/2},\]
where we have used relations (\ref{dragos1}) and (\ref{dragos2}) in the last inequality. This concludes the proof. 
\end{proof}

\begin{remark} It follows from (\ref{vortest}) that the estimate for $v^{\vare}$ in Theorem \ref{supest} is uniform in $\vare$ and $t$. 
\end{remark} 

\subsection{Harmonic vector fields and the cutoff function}

The other estimates we require in order to study the asymptotic problem involve derivatives of velocity. Before we begin examining these estimates we must address the issue that $\omega^{\vare}$ and $u^{\vare}$ are defined on an $\vare$-dependent domain, and if we want to explore their asymptotics, and, to use standard functional analysis reasoning, we must make sure they are all in the same function space. We will introduce a suitable $\vare$-dependent cutoff function for a neighborhood of $\Omega_{\vare}$ which, for reasons which will become clear
in the next section, is adapted to the geometry of the domains.

Let $\phi \in C^{\infty}(\R)$ a cutoff function with the properties that $0 \leq \phi \leq 1$, $\phi$ is monotone increasing, $\phi(s) = 1$ if $s \geq 2$ and $\phi(s) \equiv 0$ is $s \leq 1+a$, for some $0<a<1$. For $x \in \Pi_{\vare}$, set
\[\phi^{\vare} = \phi^{\vare}(x) = \phi(|T^{\vare}(x)|) = \phi(|T(x/\vare)|), \]
and $\phi^{\vare}(x) = 0$ for $x \in \overline{\Omega_{\vare}}$. Clearly $\phi^{\vare} \in C^{\infty}(\R^2)$, vanishing on a neighborhood of $\overline{\Omega_{\vare}}$. We require some properties of $\nabla\phi^{\vare}$, which we collect in the following Lemma.

\begin{lemma} \label{faieps} The cutoff $\phi^{\vare}$ defined above has the following properties:
\begin{enumerate}
\item if $H^{\vare}$ is the harmonic vector field for $\Pi_{\vare}$ then $H^{\vare} \cdot \nabla \phi^{\vare} \equiv 0$ in $\Pi_{\vare}$,
\item there exists a constant $C>0$ such that $|\nabla \phi^{\vare}| \leq C/{\vare}$,
\item there exists a constant $C>0$ such that the Lebesgue measure of the the support of $\nabla \phi^{\vare}$ is bounded by $C\vare^2$. 
\end{enumerate}
In particular, $\|\nabla \phi^{\vare}\|_{L^1(\R^2)} \to 0$ 
and $\|\nabla \phi^{\vare}\|_{L^2(\R^2)}$ is bounded 
as $\vare \to 0$.
\end{lemma}  

\begin{proof}
First we observe that
\[H^{\vare}(x) = \frac{1}{2\pi}\nabla^{\perp}\log|T^{\vare}(x)| =   \frac{1}{2\pi |T^{\vare}(x)|} \nabla^{\perp}|T^{\vare}|(x), \]
and
\[\nabla \phi^{\vare} = \phi^{\prime}(|T^{\vare}(x)|)\nabla |T^{\vare}|(x),\]
so that the first assertion follows.

Next we compute
\[|\nabla \phi^{\vare}(x)| \leq C \left| \frac{T^{\vare}(x)}{|T^{\vare}(x)|}DT^{\vare}(x)\right| \leq \frac{C}{\vare},\]
by (\ref{derivest}) and since $DT^{\vare}(x) = (1/\vare)DT(x/\vare)$.

Finally, the support of $\nabla \phi^{\vare}$ is included in the set $\{x \in \Pi_{\vare} \;|\; 1+a \leq |T^{\vare}(x)| \leq 2\}$. The Lebesgue measure of this set can be estimated by:
\[\int_{1+a \leq |T^{\vare}(x)| \leq 2} dx = \int_{1+a \leq |T(y)| \leq 2} \vare^2 dy
= C\vare^2. \]
\end{proof}

We will use the cutoff $\phi^{\vare}$ to uniformize the domains under consideration. All the functions and vector fields defined in $\Pi_{\vare}$ are hereafter to be extended arbitrarily to the full plane (for instance by assigning zero value inside $\Omega_{\vare}$). We will only be working with these extensions multiplied by $\phi^{\vare}$, which makes the extension chosen irrelevant.

We require more detailed information on the asymptotic behavior of $H^{\vare}$ than what was provided by observation (\ref{cizania}). This information is encoded in the following Lemma.

\begin{lemma} \label{Hcomp}
Let $H \equiv x^{\perp}/2\pi|x|^2$ denote the basic harmonic vector field on $\R^2 \setminus \{0\}$ and fix $R>0$. Then,
\[\phi^{\vare}H^{\vare} \rightarrow H, \]
strongly in $L^1(B(0;R))$ as $\vare \to 0$. 
\end{lemma}

\begin{proof}
We estimate directly:
\[\int_{|x| \leq R} |\phi^{\vare}H^{\vare}-H|dx \leq \int_{\{ |x|\leq R \} \cap \{|T^{\vare}(x)|\geq 2\}}|\phi^{\vare}H^{\vare}-H|dx +  
\int_{1+a \leq |T^{\vare}(x)| \leq 2}|\phi^{\vare}H^{\vare}|dx  \]
\[ +  \int_{\Omega_{\vare}\cup\{|T^{\vare}(x)| \leq 2\}} |H|dx 
\equiv {\mathcal I}_1 + {\mathcal I}_2 + {\mathcal I}_3. \]
We begin by noting that ${\mathcal I}_3 \to 0$ as $\vare \to 0$ because the function $|H|$ is locally integrable and the
Lebesgue measure of the set $\Omega_{\vare}\cup\{|T^{\vare}(x)| \leq 2\}$ tends to zero, since it is contained in a ball centered at the origin, with vanishing radius. Next we estimate ${\mathcal I}_2$:
\[|{\mathcal I}_2|  \leq  \int_{1+a \leq |T^{\vare}(x)| \leq 2}|H^{\vare}|dx = \vare \int_{1+a \leq |T(y)| \leq 2} |H_{\Pi}(y)| dy,\]
which clearly vanishes as $\vare \to 0$. Above we have changed variables to $y=x/\vare$ and we used the scaling of $H^{\vare}$ from (\ref{hvare}).

Lastly we address ${\mathcal I}_1$. We find:
\begin{align*}
|{\mathcal I}_1| 
&= \vare \int_{\{|y|\leq R/\vare \} \cap \{|T(y)|\geq 2\}}  |H_{\Pi}(y)-H(y)|dy\\
& = \vare \int_{\{|y|\leq R/\vare \}\cap \{|T(y)|\geq 2\}}
 \left|\frac{DT^t(y)T(y)^{\perp}}{2\pi|T(y)|^2} -
   \frac{y^{\perp}}{2\pi|y|^2}\right|dy. 
\end{align*}
Now we will use the expression for $T$ from Lemma \ref{confmap}, $T(y) = \beta y + h(y)$, with $\beta \in \R$, $\beta \neq 0$, $h$ a holomorphic, bounded function whose derivative $|Dh (y)| \leq C/|y|^2$ for some $C>0$. We then find:
\begin{align*} 
|{\mathcal I}_1| &= C \vare \int_{\{|y|\leq R/\vare \}\cap \{|T(y)|\geq 2\}}  \left|   
\frac{(\beta \mathbf{I} + Dh^t(y))(\beta y + h(y))^{\perp}}{|\beta y + h(y)|^2} - \beta \mathbf{I} \frac{\beta y^{\perp}}{|\beta y|^2} \right|dy \\
&\leq C \vare \int_{\{|y|\leq R/\vare \}\cap \{|T(y)|\geq 2\}}  \left|   
\frac{(Dh^t(y))(\beta y + h(y))^{\perp}}{|\beta y + h(y)|^2} \right| dy \\
& \qquad\qquad\qquad\qquad+ C \vare \int_{\{|y|\leq R/\vare \}\cap \{|T(y)|\geq 2\}} \left|
\beta \mathbf{I}\left(\frac{(\beta y + h(y))^{\perp}}{|\beta y + h(y)|^2} - 
\frac{\beta y^{\perp}}{|\beta y|^2} \right) \right|dy \\
& \leq C \vare \int_{|T(y)|\geq 2} \frac{1}{|y|^3}dy + 
C \vare \int_{\{|y|\leq R/\vare \}\cap \{|T(y)|\geq 2\}}\frac{|h(y)|}{|y||\beta y + h(y)|} dy,\\
 \intertext{using (\ref{drid}) in the second integral,} 
& \leq C (\vare + \vare \log(R/\vare)),
\end{align*}
which vanishes as $\vare \to 0$, as we wished.
\end{proof}

\subsection{Temporal estimates}

        The standard way to derive temporal estimates from spatial regularity is to use the PDE directly. In our problem this is easy to do for vorticity, i.e., system $\eeps$ and Theorem \ref{supest} imply that, for any $T>0$, $\omega^{\vare}_t$ is bounded in $L^{\infty}([0,T];W^{-1,1}_{\loc}(\R^2))$. Indeed, 
\[ \omega^{\vare}_t = - \mbox{ div }((v^{\vare} + (\alpha-m)H^{\vare})\omega^{\vare}), \]
with $\omega^{\vare}$ and $v^{\vare}$ bounded in $L^{\infty}$ and $H^{\vare}$ bounded in $L^1_{\loc}$, see (\ref{cizania}). However, we require temporal information on velocity, which is more difficult to obtain, because of the presence of the pressure in the velocity equation. Obtaining such estimates is precisely the subject of the article \cite{HLNS}. The specific case of the exterior domain was not discussed in that paper, so we will adapt the technique used there to the present situation. 

        Recall the definition of the stream function (\ref{strfnct}) and introduce the
analogous family of stream functions given by $\psi^{\vare} \equiv G^{\vare}[\omega^{\vare}].$

\begin{proposition} \label{timest}
For each $R,T>0$ there exists a constant $C=C(R,T)>0$ such that
\[ \left| \int_{\Pi_{\vare}} \varphi(x) \psi^{\vare}_t(x,t) dx \right| \leq C \|\varphi\|_{L^{\infty}}^{1/2}\|\varphi\|_{L^1}^{1/2}, \]
for every $\varphi$ in $C_0(\Pi_{\vare})$ with support contained in $B(0;R)$ and all $0\leq t\leq T$.
\end{proposition}

\begin{proof}
First we observe that there exists $R>0$ such that $\omega^{\vare}(\cdot,t)$ has support contained in the ball $B(0;R)$ for all $0 \leq t \leq T$ and $\vare > 0$. To see this, first let $R_0$ be such that $B(0;R_0)$ contains the support of $\omega_0$ and note that equation $\eeps$ means that $\omega^{\vare}$ is transported by the velocity field $u^{\vare} = v^{\vare} + (\alpha - m) H^{\vare}$, with $v^{\vare}$ uniformly bounded by a constant $C$ independent of $\vare$ and $t$ and $H^{\vare}$ is bounded by another constant $C$, also independent of $\vare$, outside of $B(0;R_0)$. Hence the support of $\omega^{\vare}(\cdot,t)$ is contained in $B(0,R_0+2Ct)$, and taking $R=R_0+2CT$ will work. Additionally, $\omega^{\vare}_t$ is also compactly supported in the same ball. We differentiate the definition of the stream function and write:
\[\psi^{\vare}_t = G^{\vare}[\omega^{\vare}_t].\]

This means that 
\[ \Delta \psi^{\vare}_t = \omega^{\vare}_t \mbox{ in } \Pi_{\vare}, \mbox{ and }
\psi^{\vare}_t = 0 \mbox{ on } \Gamma_{\vare}. \]
  
We require information on the behavior of $\psi^{\vare}_t$ near infinity, which can be readily obtained from the compactness of the support of $\omega^{\vare}_t$ and the explicit expression for $G^{\vare}$, given by (\ref{greensfctn}), substituting $T$ by $T^{\vare}$. We obtain that, for each fixed $\vare$, 
\begin{equation} \label{warny1}
\left| \psi^{\vare}_t(x,t) - L[\omega^{\vare}_t(\cdot,t)](x) \right| = {\mathcal O}(1/|x|), \mbox{ when } |x| \to \infty,
\end{equation} 
with, the functional $L$ defined by 
\[ \zeta \mapsto L[\zeta] \equiv -\frac{1}{2\pi} \int_{\Pi_{\vare}} \log|T^{\vare}(y)| \zeta(y) dy,\]
for any test function $\zeta$.

Additionally, by (\ref{inftyworry}), we have that
\begin{equation} \label{warny2}
|\nabla \psi^{\vare}_t| = {\mathcal O}(1/|x|^2), \mbox{ when } |x| \to \infty. 
\end{equation}

Let $\varphi$ be a fixed test function in $C_0(\Pi_{\vare})$ with support contained in $B(0;R)$. Let $M = \int_{\Pi_{\vare}} \varphi(x) dx$. Define 
\[\eta \equiv G^{\vare}[\varphi] + \frac{M}{2\pi}\log|T^{\vare}|. \]
Then $\eta$ satisfies the following properties:
\[ \Delta \eta = \varphi \mbox{ in } \Pi_{\vare} \mbox{, } \eta= 0 \mbox{ on } \Gamma_{\vare}, \]
\begin{equation} \label{warny3}
\eta(x) = \frac{M}{2\pi}\log|T^{\vare}(x)| + L[\varphi] + {\mathcal O}(1/|x|), \mbox{ when } |x| \to \infty, 
\end{equation}
and 
\begin{equation} \label{warny4}
|\nabla \eta| = \frac{M}{2\pi}\nabla (\log |T^{\vare}|) + {\mathcal O}(1/|x|^2), \mbox{ when } |x| \to \infty. 
\end{equation}

Then we have:
\begin{multline*}
 \left| \int_{\Pi_{\vare}} \varphi(x) \psi^{\vare}_t(x,t) dx \right| =  \left| \int_{\Pi_{\vare}} \Delta \eta(x) \psi^{\vare}_t(x,t) dx \right|= \\
= \left| \int_{\Pi_{\vare}} \eta(x) \Delta\psi^{\vare}_t(x,t) dx 
+ \int_{\partial \Pi_{\vare}} (\psi^{\vare}_t \nabla \eta  - \eta \nabla
 \psi^{\vare}_t) \cdot \hat{n} ds \right| \equiv |{\mathcal I} +
 {\mathcal J}|,
\end{multline*}
where the boundary terms include the terms at infinity.
Looking at the boundary terms expressed in ${\mathcal J}$, we see that the terms integrated on $\Gamma_{\vare}$ vanish, whereas from the asymptotic formulas (\ref{warny1} - \ref{warny4}) the terms at infinity are bounded, in such a way that we arrive at:
\begin{equation} \label{wpt1}  
|{\mathcal J}| \leq CML[\omega^{\vare}_t] \leq C\|\varphi\|_{L^1} |L[\omega^{\vare}_t]|. 
\end{equation}
 
We claim that $|L[\omega^{\vare}_t]|$ is bounded, uniformly in $0\leq t\leq T$ and $\vare>0$.
To see that we use $\eeps$ in the following way
\begin{align*}
|L[\omega^{\vare}_t]|& = \frac{1}{2\pi}\left| \int_{\Pi_{\vare}}(\log|T^{\vare}(y)|)
u^{\vare}(y,t)\cdot\nabla \omega^{\vare}(y,t)dy \right| \\
&= \frac{1}{2\pi}\left| \int_{\Pi_{\vare}}\nabla(\log|T^{\vare}(y)|)\cdot
u^{\vare}(y,t)\omega^{\vare}(y,t)dy \right| \\
&= \frac{1}{2\pi}\left|\int_{\Pi_{\vare}}\nabla(\log|T^{\vare}(y)|)\cdot
(v^{\vare}(y,t) + (\alpha-m)H^{\vare}(y))\omega^{\vare}(y,t)dy \right|.
\end{align*}

By (\ref{theH}) and (\ref{hvare}), one way of expressing the harmonic vector field $H^{\vare}$ is as
\[H^{\vare} = \frac{1}{2\pi} \nabla^{\perp}\log|T^{\vare}|, \]
so that the dangerous term in the integral above disappears, leaving:
\begin{multline*}
|L[\omega^{\vare}_t]| = \left|\int_{\Pi_{\vare}} (H^{\vare}(y))^{\perp} \cdot
v^{\vare}(y,t)\omega^{\vare}(y,t)dy \right| \leq \|H^{\vare}\|_{L^1(B(0;R))}\|v^{\vare}\omega^{\vare}\|_{L^{\infty}} \\
\leq C\|H^{\vare}\|_{L^1(B(0;R))}\|\omega_0\|_{L^{\infty}} \leq C, 
\end{multline*}
by Theorem \ref{supest}, (\ref{cizania}) and (\ref{vortest}). Together with (\ref{wpt1}), this means that 
\[|{\mathcal J}| \leq C \|\varphi\|_{L^1} \leq C \|\varphi\|_{L^{\infty}}^{1/2}
\|\varphi\|_{L^1}^{1/2},\]
since the support of $\varphi$ is contained in $B(0;R)$. We are left with estimating $|{\mathcal I}|$.

We have:
\begin{multline*}
|{\mathcal I}| = \left| \int_{\Pi_{\vare}} \eta(x) \Delta\psi^{\vare}_t(x,t) dx \right| =
\left| \int_{\Pi_{\vare}} \eta(x) u^{\vare} \cdot \nabla \omega^{\vare}
  dx \right| \\
= \left| \int_{\Pi_{\vare}} \nabla\eta(x) \cdot (v^{\vare}+ (\alpha-m)H^{\vare}) \omega^{\vare} dx \right|, 
\end{multline*}
with no boundary terms because $u^{\vare}$ is tangent to $\Gamma_{\vare}$ and $\omega^{\vare}$ has compact support. Thus,
\[|{\mathcal I}| \leq \|\nabla \eta\|_{L^{\infty}} \|(v^{\vare}+ (\alpha-m)H^{\vare}) \omega^{\vare}\|_{L^1(B(0;R))} \leq C\|\nabla \eta\|_{L^{\infty}}, \]
again using Theorem \ref{supest}, (\ref{cizania}) and (\ref{vortest}).

Finally, note that, by the definition of $\eta$, we find that
\[\nabla^{\perp}\eta = K^{\vare}[\varphi] + MH^{\vare}, \]
with $M = \int \varphi$. Thus, we use Theorem \ref{supest}, with $\varphi$ in place of $\omega^{\vare}$, to conclude that:
\[\|\nabla \eta\|_{L^{\infty}} \leq C \|\varphi\|_{L^{\infty}}^{1/2} \|\varphi\|_{L^1}^{1/2}. \]
Putting together the estimates for
$|{\mathcal I}|$ and $|{\mathcal J}|$ concludes the proof.
\end{proof}

We write the conclusion of this Proposition more explicitly in terms of temporal estimates for the quantities of interest in the Corollary below. Recall $\phi^{\vare}$ the cutoff function from the previous section and consider $\phi^{\vare}\omega^{\vare}$, $\phi^{\vare}v^{\vare}$ and $\phi^{\vare} \psi^{\vare}$ as functions (or vector fields) defined in the full plane.

\begin{corollary} \label{tempest}
Let $R, T>0$. Then there exists a constant $C=C(R,T)>0$ such that
\begin{equation} \label{vtest}
\|(\phi^{\vare} v^{\vare})_t (\cdot,t)\|_{H^{-3}(B(0;R))} \leq C,
\end{equation}
for all $\vare$ and $0 \leq t \leq T$.
\end{corollary}
\begin{proof}
Let $\zeta \in H^3_0(B(0;R)) \times H^3_0(B(0;R))$. Then, as $v^{\vare}_t = \nabla^{\perp}\psi^{\vare}_t$, 
\begin{align*} 
|\langle \zeta, (\phi^{\vare} v^{\vare})_t (\cdot,t) \rangle| 
&= \left| \int \zeta \phi^{\vare} v^{\vare}_t (\cdot,t)  \right| 
= \left| \int \mbox{ curl }(\zeta \phi^{\vare}) \psi^{\vare}_t (\cdot,t)  \right| \\
&= \left| \int \mbox{ curl }(\zeta) \phi^{\vare} \psi^{\vare}_t (\cdot,t)   + 
 \int \zeta \cdot \nabla^{\perp}\phi^{\vare} \psi^{\vare}_t (\cdot,t)  \right| \\
& \leq C \| \phi^{\vare} \mbox{ curl } \zeta \|_{L^{\infty}}^{1/2}\| \phi^{\vare} \mbox{ curl } \zeta \|_{L^{1}}^{1/2}  + C \| \zeta \nabla \phi^{\vare}\|_{L^{\infty}}^{1/2}
\|\zeta \nabla \phi^{\vare}\|_{L^1}^{1/2},\\
\intertext{by Proposition \ref{timest} used first with $\varphi = \phi^{\vare} \mbox{ curl } \zeta$ and then with 
$\varphi = \zeta \cdot \nabla^{\perp}\phi^{\vare}$,}
& \leq C(\|\mbox{ curl }\zeta \|_{L^{\infty}} +  \|\zeta \|_{L^{\infty}}),\\
\intertext{since $\|\nabla \phi^{\vare}\|_{L^{\infty}} \leq C/\vare$ and $\|\nabla \phi^{\vare} \|_{L^1} \leq C\vare$, as proved in Lemma \ref{faieps}, items 2 and 3,}
& \leq C \| \zeta \|_{H^3(B(0;R))},
\end{align*}
again by the Sobolev embedding theorem.
 This gives (\ref{vtest}).
\end{proof}

\section{Passing to the limit}

        In this section we will use the estimates developed in the previous section to obtain limit equations describing the behavior of the flow obtained in the limit $\vare \to 0$.

\subsection{Strong compactness in velocity}

        We will use a parametrized version of Tartar and Murat's Div-Curl Lemma to derive strong compactness in $L^2$ for the sequence of velocities $\phi^{\vare}v^{\vare}$. We will include here the precise statement of this version of the Div-Curl Lemma, whose proof can be found in \cite{LNT00}.

\begin{lemma} \label{tdivcurl}
Fix $T>0$ and let $\{ F^{\vare}(\cdot,t) \}$ and $\{ G^{\vare}(\cdot,t)\}$ be vector fields on $\R^n$ for 
$0\leq t \leq T$. Suppose that:
\begin{enumerate}
\item both $F^{\vare} \rightarrow F$ and $G^{\vare} \rightarrow G$
      weak-$\ast$ in $L^{\infty}([0,T];L^2_{\loc}(\R^n;\R^n))$ and   
also strongly in $C([0,T];H^{-1}_{\loc}(\R^n;\R^n))$;
\item $\{ \mbox{div } F^{\vare} \}$ is precompact in $C([0,T];H^{-1}_{\loc}(\R^n))$;
\item $\{ \mbox{curl } G^{\vare} \}$ is precompact in $C([0,T];H^{-1}_{\loc}(\R^n;\Bbb{A}^n))$.
\end{enumerate}
Then  $F^{\vare} \cdot G^{\vare} \rightharpoonup F \cdot G$ in 
${\mathcal D}^{\prime}([0,T] \times \R^n)$.
\end{lemma}

We will use the Div-Curl Lemma with $F^{\vare} = G^{\vare} = \phi^{\vare}v^{\vare}$. We begin by observing that by Theorem \ref{supest} we know that $\{\phi^{\vare}v^{\vare}\}$  is bounded in $L^{\infty}([0,T] \times \R^2)$, which is contained in $L^{\infty}([0,T];L^2_{\loc}(\R^2))$.
By Corollary \ref{tempest}, $\{(\phi^{\vare}v^{\vare})_t\}$ is bounded in $L^{\infty}([0,T];H^{-3}_{\loc})$, so that the sequence  $\{\phi^{\vare}v^{\vare}\}$ is equicontinuous from $[0,T]$ to $H^{-3}_{\loc}$. Recall that $L^2_{\loc}$ is compactly embedded into $H^{-3}_{\loc}$, so that we can use the Aubin-Lions Lemma (see \cite{temam}) to conclude that  $\{\phi^{\vare}v^{\vare}\}$ is precompact in $C([0,T];H^{-1}_{\loc})$. Passing to a subsequence if necessary, we conclude that there exists $v \in C([0,T];H^{-1}_{\loc}) \cap L^{\infty}([0,T] \times \R^2)$ such that
\begin{equation} \label{defv}
\phi^{\vare}v^{\vare} \rightarrow  v 
\end{equation}
strongly in $C([0,T];H^{-1}_{\loc})$ and weak-$\ast$ in $L^{\infty}([0,T];L^2_{\loc})$.

\begin{theorem} \label{strcomp}
We have that $\phi^{\vare}v^{\vare} \to v$ strongly in $L^2_{\loc}([0,T]\times\R^2)$.
\end{theorem}

\begin{proof}
It is enough to verify the remaining hypothesis of the Div-Curl Lemma, i.e. that 
$\mbox{div }(\phi^{\vare}v^{\vare})$ and $\mbox{curl }(\phi^{\vare}v^{\vare})$ are precompact in $C([0,T];H^{-1}_{\loc})$. First note that, by Corollary \ref{tempest}, both 
$(\mbox{div }(\phi^{\vare}v^{\vare}))_t = \mbox{div }(\phi^{\vare}v^{\vare})_t$ and $(\mbox{curl }(\phi^{\vare}v^{\vare}))_t = \mbox{curl }(\phi^{\vare}v^{\vare})_t$ are bounded in $L^{\infty}([0,T];H^{-4}_{\loc})$. Also, we have that 
\[\mbox{div }(\phi^{\vare}v^{\vare}) = v^{\vare} \cdot \nabla \phi^{\vare}, \]
which, by Lemma \ref{faieps} and Theorem \ref{supest} is bounded in $L^{\infty}([0,T];L^2)$, and 
\[\mbox{curl }(\phi^{\vare}v^{\vare}) = v^{\vare} \cdot \nabla^{\perp} \phi^{\vare} + \phi^{\vare}\omega^{\vare}, \]
which, by Lemma \ref{faieps}, Theorem \ref{supest} and (\ref{vortest}) is bounded in  
$L^{\infty}([0,T];L^2)$. Since $L^2$
is compactly embedded in $H^{-1}_{\loc}$, once again using the Aubin-Lions Lemma, we conclude that both the divergence and the curl are precompact in $C([0,T];H^{-1}_{\loc})$.
By Lemma \ref{tdivcurl}, we infer that 
\[ |\phi^{\vare}v^{\vare}|^2 \rightharpoonup |v|^2 \mbox{ in } {\mathcal D}^{\prime}, \]
which in turn implies that $\phi^{\vare}v^{\vare}$ converges strongly to $v$ in $L^2_{\loc}$,  
as we wished.
\end{proof}

\subsection{The asymptotic vorticity equation}

We begin by observing that the sequence $\{\phi^{\vare}\omega^{\vare}\}$ is bounded in $L^{\infty}([0,T]\times\R^2)$, so that, passing to a subsequence if necessary, we have that 
\begin{equation} \label{vortconv}
\phi^{\vare}\omega^{\vare} \rightharpoonup \omega, \mbox{ weak}\ast L^{\infty}([0,T]\times\R^2). \end{equation}

We are already in possession of a limit velocity 
\begin{equation} \label{theU}
u=v+(\alpha-m)H,
\end{equation}
where $v$ is the strong limit of $\phi^{\vare}v^{\vare}$ from the previous subsection and $H$ is the strong limit of $\phi^{\vare}H^{\vare}$, introduced in Lemma \ref{Hcomp}. The purpose of this section is to prove that $\omega$ and $u$ satisfy, in an appropriate weak sense, the system:

\begin{equation} \label{limitvorteq} 
\left\{ \begin{array}{ll}
\omega_t + u \cdot \nabla \omega = 0, & \mbox{ in } \R^2 \times (0,\infty)  \\
\mbox{div }u=0 \mbox{ and } \mbox{ curl }u = \omega + (\alpha-m)\delta, & \mbox{ in } \R^2 \times [0,\infty)   \\
|u| \rightarrow 0, &  \mbox{ as } |x| \rightarrow \infty \\
\omega(x,0) = \omega_0(x), & \mbox{ in } \R^2. 
\end{array} \right. 
\end{equation}

Above, $\delta$ is the Dirac delta centered at the origin.

\begin{definition} \label{wksollim}
The pair $(u,\omega)$, with $u$ a vector field in $L^{\infty}([0,\infty);L^1_{\loc}(\R^2))$ and $\omega \in L^{\infty}([0,\infty);L^{\infty}(\R^2))$ is a weak solution of system (\ref{limitvorteq}) if
\begin{enumerate}
\item For any test function $\varphi \in C^{\infty}_c([0,\infty)\times\R^2)$ we have:
\[\int_0^{\infty} \int_{\R^2} \varphi_t \omega\, dxdt + 
\int_0^{\infty} \int_{\R^2}\nabla \varphi \cdot u \omega\, dxdt +
\int_{\R^2} \varphi(x,0)\omega_0(x)\, dx = 0, \]
\item we have $\mbox{ div }u = 0$ and $\mbox{ curl }u = \omega + (\alpha -m)\delta$ in the sense of distributions, with $|u| \to 0$ at infinity.
\end{enumerate} 
\end{definition}

When $\alpha = m$ the definition above reduces to the standard definition of weak solution for the vorticity formulation of the incompressible 2D Euler equations. We now state and prove the main result in this article.

\begin{theorem} \label{mainth}
The pair $(u,\omega)$ given by (\ref{theU}) and (\ref{vortconv}) is a weak solution of the system (\ref{limitvorteq}). 
\end{theorem}  

\begin{proof}
We will begin by verifying that $(u,\omega)$ satisfy the linear elliptic system corresponding to the second condition in Definition \ref{wksollim}.
Recall that $u^{\vare} = v^{\vare} + (\alpha - m)H^{\vare}$. First observe that $\phi^{\vare}u^{\vare} \to u$ strongly in $L^1_{\loc}$, by Lemma \ref{Hcomp} and Theorem \ref{strcomp}. Hence, by Lemma \ref{faieps} and Theorem \ref{supest}, 
\[\mbox{div }u = \lim_{\vare\to 0} \mbox{ div }(\phi^{\vare} u^{\vare}) =
\lim_{\vare\to 0} v^{\vare} \cdot \nabla \phi^{\vare} = 0, \]
where the limits were taken in the sense of distributions. Similarly, we use that $\phi^{\vare} v^{\vare} \to v$ strongly in $L^2_{\loc}$, so that 
\[\mbox{curl }v = \lim_{\vare\to 0} \mbox{ curl }(\phi^{\vare} v^{\vare}) =
\lim_{\vare\to 0} \phi^{\vare}\omega^{\vare} +  v^{\vare} \cdot \nabla^{\perp} \phi^{\vare} = \omega, \]
in the sense of distributions. Hence, 
\[\mbox{curl }u = \mbox{curl }v + (\alpha-m)\mbox{curl }H = \omega + (\alpha-m)\delta. \]

The velocity $u$ satisfies the condition $|u| \to 0$ at infinity because the convergence of $\phi^{\vare}u^{\vare}$ to $u$ is uniform outside a ball containing the origin, as can be checked directly by the explicit expressions for $K^{\vare}[\omega^{\vare}]$ and $H^{\vare}$, using the uniform compact support of $\omega^{\vare}$.

Next, we introduce an auxiliary nonlinear functional $I_{\vare}$. Given any test function $\varphi \in C^{\infty}_c([0,\infty)\times\R^2)$ let:
\[I_{\vare}[\varphi] \equiv \int_0^{\infty} \int_{\R^2} \varphi_t (\phi^{\vare})^2 \omega^{\vare} \, dxdt + 
\int_0^{\infty} \int_{\R^2} \nabla \varphi \cdot (\phi^{\vare} u^{\vare}) (\phi^{\vare} \omega^{\vare}) \, dxdt. \]

Fix $\varphi \in C^{\infty}_c([0,\infty)\times\R^2)$. The proof that $(u,\omega)$ is a weak solution proceeds in two steps. First we will show that 
\[I_{\vare}[\varphi] + \int_{\R^2} \varphi(x,0)\omega_0(x) \, dx \to 0, \]
when $\vare \to 0$. The second step consists of showing that 
\[I_{\vare}[\varphi] \to \int_0^{\infty} \int_{\R^2} \varphi_t \omega\, dxdt + \int_0^{\infty} \int_{\R^2}\nabla \varphi \cdot u \omega\, dxdt, \]
as $\vare \to 0$. Clearly these two steps complete the proof. 

We address the first step. As $u^{\vare}$ and $\omega^{\vare}$ satisfy $\eeps$, it can be easily seen that
\begin{multline*}
\int_0^{\infty} \int_{\R^2} \varphi_t (\phi^{\vare})^2 \omega^{\vare} \, dxdt  \\
= -\int_0^{\infty}\int_{\R^2} \nabla(\varphi (\phi^{\vare})^2) \cdot
u^{\vare} \omega^{\vare} \, dxdt - \int_{\R^2}
\varphi(x,0)(\phi^{\vare})^2(x) \omega_0(x) \, dx. 
\end{multline*}
Thus we compute:
\begin{align*}
I_{\vare}[\varphi] 
&= -2 \int_0^{\infty}\int_{\R^2} \varphi \nabla\phi^{\vare} \cdot (\phi^{\vare}u^{\vare}) \omega^{\vare} \, dxdt - 
\int_{\R^2} \varphi(x,0)(\phi^{\vare})^2(x) \omega_0(x) \, dx \\
& = -2 \int_0^{\infty}\int_{\R^2} \varphi \nabla\phi^{\vare} \cdot (\phi^{\vare}v^{\vare}) \omega^{\vare} \, dxdt - 
\int_{\R^2} \varphi(x,0)(\phi^{\vare})^2(x) \omega_0(x) \, dx, 
\end{align*}
since $H^{\vare} \cdot \nabla \phi^{\vare} = 0$, by Lemma \ref{faieps}. It is easily seen that
\[\left| I_{\vare}[\varphi] + \int_{\R^2} \varphi(x,0)(\phi^{\vare})^2(x) \omega_0(x) \, dx \right| \leq \|\omega^{\vare}\|_{L^{\infty}} \|\phi^{\vare}v^{\vare}\|_{L^{\infty}} \|\varphi \nabla \phi^{\vare}\|_{L^1} \to 0,\]
as $\vare \to 0$, by (\ref{vortest}), Theorem \ref{supest} and Lemma \ref{faieps}. On the other hand, since the support of $\omega_0$ does not contain the origin, it follows that for all $\vare$ sufficiently small, $(\phi^{\vare})^2 \omega_0 = \omega_0$. This concludes the first step.

For the second step we begin by noting that the linear term offers no difficulty. The nonlinear term consists of the weak-strong pair vorticity-velocity, with the vorticity $\phi^{\vare} \omega^{\vare}$ converging in the weak-$\ast$ topology of $L^{\infty}((0,\infty)\times\R^2)$ to $\omega$ and the localized velocity $\nabla \varphi \cdot (\phi^{\vare} u^{\vare})$ converging in the strong topology of $L^1((0,\infty)\times \R^2)$ to $u$. This concludes the second step.
\end{proof}
 
If $\alpha = m$ the whole sequence $\phi^{\vare} u^{\vare}$ converges
to $u$, without needing to pass to a subsequence, due to the
uniqueness portion of Yudovich's Theorem, see \cite{yudovich63}. Indeed, our argument shows that for any sequence $\omega_k \to 0$, there exists a subsequence $\omega_{k_j}$ such that $\omega^{\vare_{k_j}} \rightharpoonup \omega$ and that $\omega$ is a weak solution of 2D Euler in the full plane with $\omega_0$ as initial data. By Yudovich's Theorem, $\omega$ is uniquely determined, so that any accumulation point of the precompact sequence $\omega^{\vare_k}$ is precisely equal to $\omega$. This implies that the whole sequence $\omega^{\vare_k}$ converges to $\omega$, and, as this subsequence was arbitrary, our contention follows. 

If $\alpha \neq m$, Theorem \ref{mainth} implies existence of a weak
solution for the limit equation (\ref{limitvorteq}), which is, roughly
speaking, the usual Euler equation with an embedded point vortex
background. Also, its restriction to $\R^2\setminus\{0\}$ is the
standard Euler equation (see relation \eqref{dieq}). The more physically meaningful presentation of the incompressible 2D Euler equations is the velocity form. Clearly, if $\alpha = m$, the limit flow satisfies the usual velocity form of the incompressible Euler equations, but if $\alpha \neq m$ it is not entirely clear what form the asymptotic balance of momentum equations should take. We clarify this issue in the next subsection.

\subsection{The asymptotic velocity equation}

	We have obtained a satisfactory description of the small obstacle limit expressed in terms of vortex dynamics. The purpose of this last subsection is to obtain a description of the limit behavior expressed in terms of flow velocity. We will avoid introducing explicitly the weak forms of the velocity equation, keeping the discussion less technical than before.  More specifically, our purpose is to determine the specific form of the momentum equations in the small obstacle asymptotics, in the case $\alpha \neq m$. The main difficulty is making sense of the term $H \otimes H$. We will present several equivalent forms for the limit equation
\eqref{limitvorteq}. Recall that 
\begin{equation*}
  H=\frac{x^\perp}{2\pi|x|^2}.
\end{equation*}

First we note that (\ref{limitvorteq}) can be re-written as:
\begin{equation}
\label{di1}
\left\{
   \begin{aligned}
\omega_t+\dive(v\omega)+(\al-m)\dive(H\omega) & =0\\
v & =K[\omega]\\
\omega(x,0) & =\omega_0(x)     
   \end{aligned}
\right.
\end{equation} 
which is clearly well defined in the sense of distributions. This can be easily seen to be equivalent to (\ref{limitvorteq}) since one has that $\dive H=0$ and $\curl H=\delta$ in
the sense of $\mathcal{D}'(\R^2)$. Writing (\ref{limitvorteq}) in this way will lead to a weak
formulation for the velocity equation. More precisely, we next show
that if \eqref{di1} holds, then we have the following equation for $v$
\begin{equation}
  \label{di2}
 \left\{
   \begin{aligned}
v_t+v\cdot\nabla v+(\al-m)\dive(v\otimes H+H\otimes v)
-(\al-m)v(0)^\perp\delta&=\nabla p\\
\dive v&=0\\
v(x,0)&=K[\omega_0].
   \end{aligned}
\right.
\end{equation}
In order to prove the equivalence of \eqref{di1} and \eqref{di2} it is
sufficient to show that
\begin{equation}
  \label{di3}
\curl\bigl[\dive(v\otimes H+H\otimes v)-v(0)^\perp\delta\bigr]
=\dive(H\curl v)  
\end{equation}
for all divergence free vector fields $v$ belonging to the space
$W^{1,p}_\loc$ for some $p>2$. Indeed, if \eqref{di3} holds then we get
for $\omega=\curl v$
\begin{align*}
0&=\curl\nabla p=\curl\bigl[v_t+v\cdot\nabla v+(\al-m)\dive(v\otimes H+H\otimes v)
-(\al-m)v(0)^\perp\delta \bigr]\\  
&=\curl(v_t+v\cdot\nabla v)
+(\al-m)\curl\bigl[\dive(v\otimes H+H\otimes v)
-v(0)^\perp\delta \bigr]  \\
&=\omega_t+v\cdot\nabla\omega+(\al-m)\dive(H\omega)
\end{align*}
so relation \eqref{di1} holds true. And vice versa, if \eqref{di1}
holds then we deduce that the left hand side of \eqref{di2} has zero
curl so it must be a gradient.

\medskip

We now prove \eqref{di3} under the hypothesis that $\dive v=0$ and
$v\in W^{1,p}_\loc$, $p>2$. First note that $H\curl v$ is well defined
since $\curl v\in L^p_\loc$ and $H\in L^q_\loc$ for all
$q<2$. Moreover, since $W^{1,p}_\loc\subset \mathcal{C}^0$, $v(0)$ is
well defined, too. Next, it suffices to prove \eqref{di3} for smooth
$v$ since we can pass to the limit on a sequence of smooth
approximations of $v$ that converge strongly in $W^{1,p}_\loc$ and
$\mathcal{C}^0$. Now, it is trivial to check that, for a $2\times 2$
matrix $A$ with coefficients distributions, the following identity
holds
\begin{equation*}
  \curl\dive A=\dive
\begin{pmatrix}
  \curl C_1\\ \curl C_2
\end{pmatrix}
\end{equation*}
where $C_i$ denotes the $i$-th column of $A$. For smooth $v$, we now
deduce that
\begin{align*}
\curl\dive(v\otimes H+H\otimes v)
&=\dive
\begin{pmatrix}
  \curl(vH_1)+\curl(Hv_1)\\
\curl(vH_2)+\curl(Hv_2)
\end{pmatrix}\\
&=\dive\bigl(H\curl v+v\cdot\nabla^\perp H 
+v\curl H+H\cdot\nabla^\perp v\bigr). 
\end{align*}
It is a simple computation to check that
\begin{equation*}
\dive(v\cdot\nabla^\perp H + H\cdot\nabla^\perp v)
= v\cdot\nabla^\perp \dive H + H\cdot\nabla^\perp \dive v
+\curl v\dive H +\curl H\dive v.
\end{equation*}

We therefore get the following general formula
\begin{multline*}
\curl\dive(v\otimes H+H\otimes v)
= \dive(H\curl v +v\curl H)
+v\cdot\nabla^\perp \dive H\\
 + H\cdot\nabla^\perp \dive v
+\curl v\dive H +\curl H\dive v. 
\end{multline*}

Taking into account that $\dive v=\dive H=0$ and $\curl H=\delta$ we
infer that
\begin{align*}
\curl\dive(v\otimes H+H\otimes v)
&= \dive(H\curl v +v\delta)\\
&=\dive(H\curl v) +\dive\bigl[v(0)\delta\bigr] \\ 
&=\dive(H\curl v) +\curl\bigl[v(0)^\perp\delta\bigr].  
\end{align*}
Relation \eqref{di3} now follows and so does the formulation
\eqref{di2}. 

\bigskip

We would now like to give a formulation in terms of $u$ only. Simply
replacing $v$ by $u-(\al-m)H$ in \eqref{di2} is not very
enlightening. On the other hand, $u$ is not $L^2_\loc$ since $H\not\in
L^2_\loc$. Therefore, a formulation which makes use of $u\otimes u$
cannot be made rigorous. Nevertheless, it is still desirable to obtain such
a formulation in order to clarify which form the limit equation for the
velocity takes. 

We will proceed as follows. First note that 
\begin{equation*}
  u\otimes u=v\otimes v+(\al-m)(v\otimes H+H\otimes v)
+(\al-m)^2H\otimes H.
\end{equation*}
All these terms except $H\otimes H$ are well defined. In order to give
a sense to $u\otimes u$ up to 0 we will simply extend $H\otimes H$ up
to 0 by its finite part $\pf(H\otimes H)$ that we define as follows.
\begin{definition}\label{defpf}
Let $g$ be a function which is homogeneous of degree $-2$ and of class
$\mathcal{C}^\infty$ on $\R^2\setminus\{0\}$. The finite part of $g$
is the following distribution on $\R^2$:
\begin{equation*}
\mathcal{C}_c^\infty(\R^2)\ni\varphi\longmapsto\scal{\pf g}\varphi
=\lim_{\ep\to0}\Bigl(\int\limits_{|x|>\ep}g\varphi-\varphi(0)\int\limits_{\ep<|x|<\la}g\Bigr).  
\end{equation*}
\end{definition}
\begin{remark}
This definition depends on the choice of the positive parameter $\la$ that we fix
once and for all.  
\end{remark}
 
We claim that if we extend $u\otimes u$ to a distribution of
$\mathcal{D}'(\R^2)$ by
\begin{equation*}
u\otimes u=v\otimes v+(\al-m)(v\otimes H+H\otimes v)
+(\al-m)^2\pf(H\otimes H)  
\end{equation*}
then the limit velocity $u$ verifies the following PDE
\begin{equation}\label{dieq}
\left\{
\begin{aligned}
u_t+\dive(u\otimes u)&=-\nabla p+(\al-m)v(0)^\perp\delta,\\
 \dive u&=0\\
u(x,0)&=K[\omega_0]+(\al-m)H.
\end{aligned}
\right. 
\end{equation}
This clearly follows from \eqref{di2} if we are able
to prove that $\dive\pf(H\otimes H)$ is a gradient. We now show that 
$\curl\dive\pf(H\otimes H)=0$ in $\mathcal{D}'(\R^2)$; this clearly
implies that $\dive\pf(H\otimes H)$ is a gradient. 
Let $\varphi\in\mathcal{C}_c^\infty(\R^2)$ be a test function. By
definition
\begin{multline*}
  \sca{\curl\dive\pf(H\otimes H)}\varphi
=-\sca{\pf(H\otimes H)}{\nabla\otimes\nabla^\perp \varphi}\\
=-\lim_{\ep\to0}\Bigl(\int\limits_{|x|>\ep}(H\otimes H)\cdot(\nabla\otimes\nabla^\perp \varphi)-\nabla\otimes\nabla^\perp \varphi(0)\cdot\int\limits_{\ep<|x|<\la}H\otimes H\Bigr).
\end{multline*}
A simple calculation shows that $\int\limits_{\ep<|x|<\la}H\otimes H$ is
a matrix proportional to the identity while $\nabla\otimes\nabla^\perp
\varphi(0)$ is a trace free matrix. Therefore 
\begin{equation*}
\nabla\otimes\nabla^\perp \varphi(0)\cdot\int\limits_{\ep<|x|<\la}H\otimes H=0.  
\end{equation*}

It is easy to check that, for $x\neq0$, 
\begin{equation*}
\dive(H\otimes H)=\frac{x}{4\pi^2|x|^4}=\nabla\bigl(\frac1{8\pi ^2|x|^2}\bigr).  
\end{equation*}

Next, we integrate by parts and use Stokes formula to deduce that
\begin{align*}
  \sca{\curl\dive\pf(H\otimes H)}\varphi
&=\lim_{\ep\to0}\Bigl(\int\limits_{|x|>\ep}\dive(H\otimes
H)\cdot\nabla^\perp \varphi 
+\frac1\ep\int\limits_{|x|=\ep}(H\otimes
H)\cdot(x\otimes\nabla^\perp \varphi)\Bigr)\\
&=\lim_{\ep\to0}\Bigl(-\frac1\ep\int\limits_{|x|=\ep}\dive(H\otimes
H)\cdot x^\perp \varphi 
+\frac1\ep\int\limits_{|x|=\ep}(H\otimes
H)\cdot(x\otimes\nabla^\perp \varphi)\Bigr),  
\end{align*}
where we have used that, for $x\neq0$, $\curl\dive(H\otimes H)=0$. On
the right-hand side, the first term vanishes since $\dive(H\otimes H)$
is proportional to $x$, that is orthogonal to $x^\perp$. The second term
also vanishes since 
\begin{equation*}
(H\otimes H)\cdot(x\otimes\nabla^\perp \varphi) 
=(H\cdot x)(H\cdot \nabla^\perp \varphi) 
=\frac1{2\pi|x|^2}(x^\perp\cdot x)(H\cdot \nabla^\perp \varphi)  
=0.
\end{equation*}

This concludes the proof of the relation $\curl\dive\pf(H\otimes H)=0$. Finally, we observe in passing that
\begin{equation*}
\dive\pf(H\otimes H)
=\frac1{8\pi^2}\nabla\bigl(\pf\frac1{|x|^2}+\pi\delta\bigr).   
\end{equation*}

\section{Final remarks and conclusion}

        The results we have proved here are not very surprising after the nature of the harmonic part of the flow has been clarified. In fact, a good illustration of the results obtained can be explicitly computed if one considers the limit flow associated to the motion of a single point vortex in the exterior of a disk of vanishing radius. To be precise, consider the motion of a single point vortex of strength $m>0$ in the exterior of the disk $B(0;r)$, $r<1$, initially located at $(1,0)$. Let $P_r = P_r(t)$ denote the trajectory of this point vortex, which will remain on the circle of radius $1$ around the origin, moving with constant angular velocity. At each fixed time the velocity field can be computed using the method of images as:
\[u_r(x,t) = \frac{m}{2\pi} \frac{(x-P_r(t))^{\perp}}{|x-P_r(t)|^2} - 
\frac{m}{2\pi} \frac{(x-r^2P_r(t))^{\perp}}{|x-r^2P_r(t)|^2} + \frac{\alpha}{2\pi}\frac{x^{\perp}}{|x|^2}.\]
This flow is equivalent to flow in the full plane associated with three point vortices: the original one at $P_r(t)$, the image point vortex at $r^2P_r(t)$, with strength $-m$ and a point vortex of arbitrary strength $\alpha$ at the center of disk. The location of the image vortex is the inversion, with respect to the disk of radius $r$, of the location of the original vortex. 
The point vortex of arbitrary strength at the center is associated with the choice of the harmonic part of the flow. Curiously, when the method of images for flow in the exterior of a disk was discussed in \cite{saffman95}, Saffman simply assumed that $\alpha = m$ was the correct choice of harmonic part. In the case $\alpha = m$, we have that
\[P_r(t) = \left(\cos\left( \frac{m r^2t}{2\pi(1-r^2)}\right), - \sin \left( \frac{mr^2t}{2\pi(1-r^2)}\right) \right),\]
so it is easy to verify that the angular velocity of $P_r(t)$ vanishes as the radius of the disk vanishes, so that $P_r(t)$ converges pointwise in time to $P_r(0)=(1,0)$. Furthermore it can be readily checked that the velocity fields $u_r$ converge to the velocity field due to a single point vortex in full plane flow as $r \to 0$ if and only if $\alpha = m$. In short, what we accomplished in this paper is to verify that the evanescent obstacle exterior flow, for a general domain geometry and general vorticity, behaves exactly like the corresponding limit of  point vortex flow in the exterior of a vanishing disk.  

\vspace{1cm}
\centereps{3in}{3in}{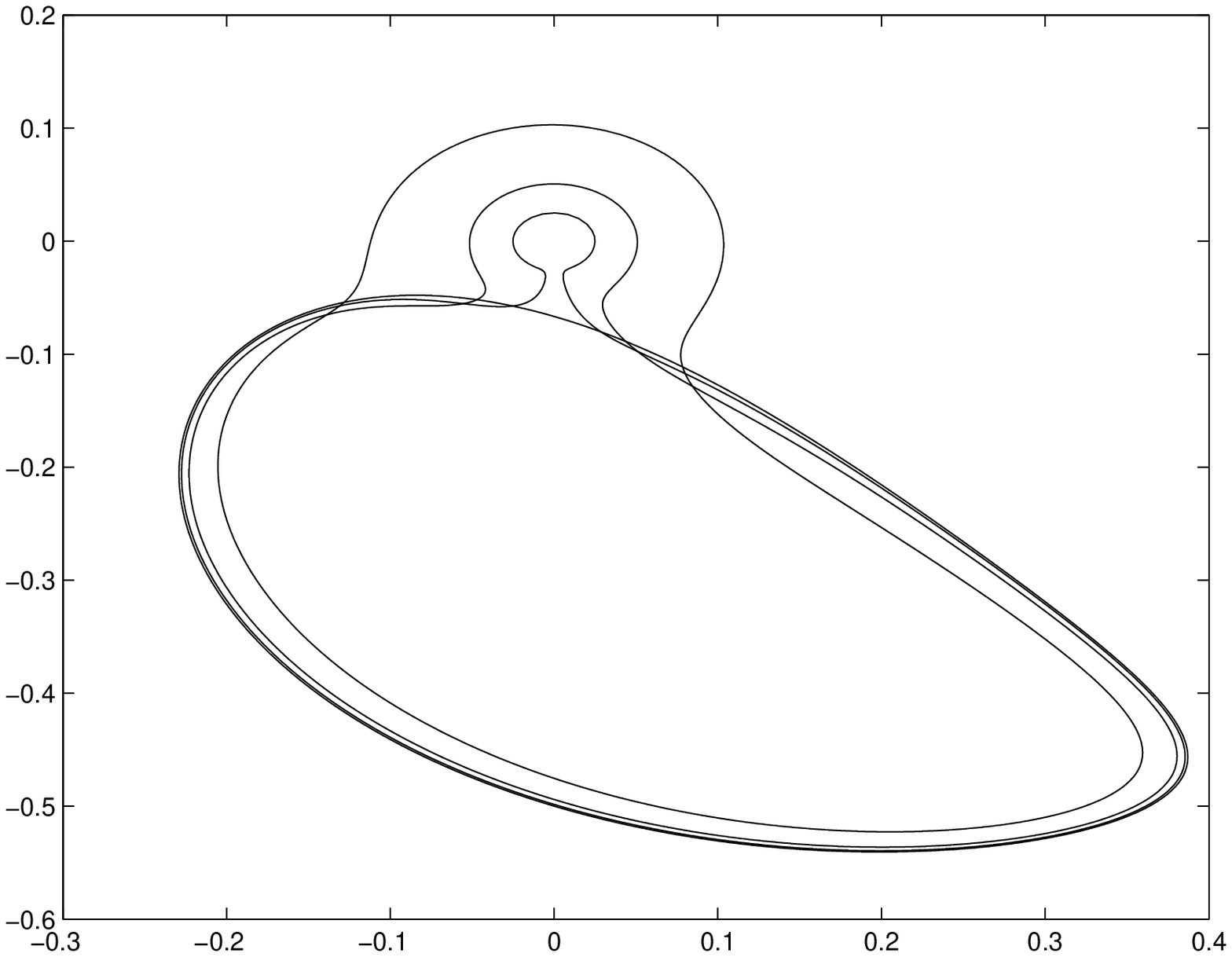}
\centerline{ {\scriptsize \em{Fig. 1 Curves $X_r(\Sigma,T)$ for $T=2$ and $r=0$; $0.025$; $0.05$; $0.1$.}}}
\vspace{1cm}

        One interesting feature of the limit process in the case $\alpha = m$ is that, although both the approximating and the limit flows are smooth, the particle trajectories do not converge uniformly. To see that, we use the illustration described above. Set $m=1=\alpha$ and let $X_r = X_r(Q,t)$ denote the particle trajectory under the flow $u_r$, starting at the Lagrangian marker $Q$, and let $X_0=X_0(Q,t)$ denote the particle trajectory under the limit flow. Let $\Sigma = \{|x| = 1/4\}$ and assume $r<1/4$. Clearly, for any $t>0$, $X_r(\Sigma,t)$ and $X_0(\Sigma,t)$ are smooth Jordan curves in the plane. It can be checked that there exists $T>0$ such that the origin lies outside $X_0(\Sigma,T)$. Fix such a time $T>0$. On one hand one expects to have  $X_r(\Sigma,T)$ converging to $X_0(\Sigma,T)$ as $r \to 0$. On the other hand, the origin lies in the interior of $X_r(\Sigma,T)$ for all $r>0$, as the flow has to remain in the exterior of $B(0;r)$, see Figure 1. This means that the convergence as $r \to 0$ of the maps $Q \mapsto X_r(Q,T)$ to the limit map $X_0(Q,T)$ cannot be uniform, even in the compact parts of their common domain.

        Below we add two additional remarks concerning the results we have obtained.

\begin{enumerate}

\item In the case $\gamma = \alpha - m \neq 0$, one may ask why fix the circulation $\gamma$ around the small obstacle, since this implies very large, maybe unphysical velocities at the boundary of the obstacle, and not consider it some appropriate function of $\vare$, perhaps vanishing when $\vare \to 0$. In fact, $\gamma$ plays the role of a passive parameter in the argument we have presented, so there would be no change in the argument if we consider $\gamma$ to be a function of $\vare$. The limit flow would depend on $\gamma(\vare)$ only through its limit when $\vare \to 0$ in precisely the same manner as presented.

\item For the sake of simplicity, we have presented our argument for smooth, compactly supported initial vorticities, but the argument can be easily performed for compactly supported vorticities in $L^p$, $p>2$. The argument does not work for $p \leq 2$ because there are serious difficulties in making sense of the term $H \omega$ in that case, and the value $v(0)$, which appears in the limit velocity (\ref{dieq}), also becomes ambiguous. The case $p \leq 2$ is thus an interesting open problem. 

\end{enumerate}

Let us point out some of the natural questions raised by the research presented here. First, an analogous question can be asked with regard to the 2D incompressible Navier-Stokes equations, and this is the subject of current investigation by the authors. For incompressible 3D Euler, the problem initially becomes proving that a smooth solution exists for a time that is independent of the size of the domain, something we did not investigate, but that appears to be difficult. Of course, the most interesting situation is the same limit for 3D Navier-Stokes, but it makes sense to work out the 2D equations first. Once the viscous problem has been understood, one may ask about the interaction of the small obstacle and small viscosity limits, which in two dimensions is a simplification of the classical open problem of convergence of Navier-Stokes solutions to Euler solutions in the presence of boundaries, see \cite{tw02} for an account of the state-of-the-art of this problem. Another natural question is to understand the same limit with more than one obstacle, with perhaps only some of the obstacles vanishing.  Finally, it would be interesting to obtain a description of the leading order correction associated to the evanescent obstacle, in the case $\alpha = m$, with respect to the unperturbed full plane problem. 

\vspace{0.8cm}

{\footnotesize  {\it Acknowledgment:} 
This research has been supported in part by the UNICAMP Differential Equations PRONEX, FAPESP grant \# 00/02097-1 and FAEP grant \# 0285/01. The authors would like to thank Prof. Paulo Cordaro, for calling our attention to the reference \cite{BK85}. We would also like to thank the generous hospitality of the Univ. de Rennes I and of the Institute of Mathematical Sciences of the Chinese University of Hong Kong.

\vskip\baselineskip

\noindent
{\sc
Drago\c s Iftimie\\
IRMAR, Universit{\'e} de Rennes I, \\
Campus de Beaulieu, 35042 Rennes, France
\\}
{\it E-mail address:} iftimie@maths.univ-rennes1.fr

\vspace{.1in}
\noindent
{\sc
Milton C. Lopes Filho\\
Departamento de Matematica, IMECC-UNICAMP.\\
Caixa Postal 6065, Campinas, SP 13083-970, Brasil
\\}
{\it E-mail address:} mlopes@ime.unicamp.br

\vspace{.1in}
\noindent
{\sc 
Helena J. Nussenzveig Lopes\\
Departamento de Matematica, IMECC-UNICAMP.\\
Caixa Postal 6065, Campinas, SP 13083-970, Brasil
\\}
{\it E-mail address:} hlopes@ime.unicamp.br

\end{document}